%% file: allpay.tex
\documentclass[11pt]{article}
\usepackage[margin=1in]{geometry}
\usepackage{amsmath,amsthm,amssymb}
\usepackage{graphicx}
\usepackage[ruled,vlined]{algorithm2e}
\usepackage{paralist}

\theoremstyle{plain} 
\newtheorem{theorem}{Theorem}[section]
\newtheorem{corollary}[theorem]{Corollary}
\newtheorem{lemma}[theorem]{Lemma}
\newtheorem{proposition}[theorem]{Proposition}
\newtheorem{definition}[theorem]{Definition}

\theoremstyle{definition}

\usepackage[backref=page]{hyperref}
\hypersetup{
  colorlinks=true,
  citecolor=blue
}

\renewcommand{\vec}[1]{\ensuremath{\mathbf{#1}}}
\def \B {\mathbf{B}}
\def \p {\mathbf{p}}
\def \q {\mathbf{q}}
\def \b {\mathbf{b}}
\def \pr {\mathrm{Pr}}

\def \b {\mathbf{b}}
\def \X {\mathbf{X}}
\def \x {\vec{x}}

\def \v {\mathbf{v}}
\def \B {\mathbf{B}}

\def \O {\mathbf{O}}
\def \o {\vec{o}}

\def \G {\mathbf{G}}
\def \g {\mathbf{g}}
\def \f {\mathbf{f}}

\DeclareMathOperator*{\E}{\mathbb{E}}
\newcommand{\down}[1]{\ensuremath{\lfloor #1 \rfloor}}
\newcommand{\up}[1]{\ensuremath{\lceil #1 \rceil}}

\title{On the Efficiency of All-Pay Mechanisms}
\author{George Christodoulou\thanks{Department of Informatics,
    University of Liverpool, UK.  Email:
    \texttt{G.Christodoulou@liverpool.ac.uk} } 
  \and Alkmini Sgouritsa\thanks{Department of Informatics, University
    of Liverpool, UK. Email: \texttt{a.sgouritsa@liverpool.ac.uk}} \and
  Bo Tang\thanks{Department of Informatics, University of Liverpool,
    UK. Email: \texttt{Bo.Tang@liverpool.ac.uk}}}
\begin{document}

\maketitle

\input{abstract}
\input{introduction}

\input{relatedWork}

\input{preliminaries}
\input{xos}
\input{multi-unit}
\input{singleItemPoA}

%\input{conclusion}

\bibliographystyle{plain}\bibliography{poa}

\begin{appendix}
  \input{appendixXosInequality}

  \input{appendixSingleFirst}
\end{appendix}

\end{document}

%% file: abstract.tex
\begin{abstract}

  % All-pay auctions are widely used to model economic agents making
  % irreversible investments in competitions. Specifically, both winners
  % and losers have to pay their bids in (first-price) all-pay
  % auctions.
  We study the inefficiency of mixed equilibria, expressed as the price of anarchy, of all-pay auctions in
  three different environments: combinatorial, multi-unit and single-item
  auctions. First, we consider item-bidding combinatorial auctions where
  $m$ all-pay auctions run in parallel, one for each good. % We
  % strengthen the best known upper bound $2$ \cite{ST13} to $1.82$ by
  % proving some structural properties that characterize the mixed Nash
  % equilibria of the game.  
  For fractionally subadditive valuations, we strengthen the upper bound from $2$
  \cite{ST13} to $1.82$ by proving some structural properties that
  characterize the mixed Nash equilibria of the game.
% by proving some particular
%   theorems that describe mixed equilibria. % best upper bound was $2$ by Syrgkanis and Tardos \cite{ST13}. Here,
  % we 
  % In order to show this improved bound we combine ideas from the
  % upper bound techniques used by Christodoulou et
  % al.~\cite{CKST13}, together with some novel structural
  % theorems that describe equilibria in simultaneous all-pay
  % auctions.
  % For combinatorial auctions, we show the mixed PoA in simultaneous
  % all-pay auctions is at most $1.82$ that improves the current bound
  % $2$ by Syrgkanis and Tardos \cite{ST13}. For multi-unit
  % auctions, we design an all-pay auction 
  % different good-allocation mechanisms, that share a common feature:
  % all participants always pay their bids regardless if they win the
  % good(s) or not. Our focus is on the analysis of their equilibrium
  % (in)efficiency expressed as the Price of Anarchy.
  Next, we design an all-pay mechanism with a randomized allocation
  rule for the multi-unit auction. We show that, for bidders with
  submodular valuations, the mechanism admits a unique, $75\%$
  efficient, pure Nash equilibrium. The efficiency of this mechanism
  outperforms all the known bounds on the price of anarchy of mechanisms used for
  multi-unit auctions. % The mechanism we
  % suggest here asks each bidder for a single bid and decides the
  % randomized allocation by running the proportional-share allocation
  % mechanism (used for divisible goods) on appropriate concave
  % valuation functions.
  Finally, we analyze single-item all-pay auctions motivated by their
  connection to contests and show tight bounds on the price of anarchy of social
  welfare, revenue and maximum bid.

% In a contest, the objective is to design a
%   reward allocation rule to maximize social welfare, sum of bids
%   (revenue) or maximum bid. For the social welfare, we show a tight
%   bound on the price of anarchy of approximately $1.185$.   
% 	For the revenue and maximum bid, we show
%   that they are at least as high as half of the second highest
%   valuation in any mixed Nash equilibrium. In contrast, when using any
%   reward structure other than allocating the entire reward to the
%   highest bidder, the revenue and maximum bid in some mixed Nash
%   equilibrium may be strictly less than half of the second highest
%   valuation.

% The Price of Anarchy is an increasing function of
% the number of players. For $2$ players it is exactly $8/7\approx 1.14$
% while for many players it is approximately .

%For the case of bidders with OXS valuations, we show that the Price of
%Anarchy of all-pay auctions is at least $e/(e-1)\approx 1.58$, 

%tight bounds, using the characterization of Baye et
% al.~\cite{BKV96}. The price of anarchy is an increasing function of
% the number of players. For 2 players it is exactly $8/7\approx 1.14$
% while for many players it is approximately $1.185$.

% We find remarkable the fact that the Price of Anarchy of all-pay
% auctions is exactly the same with the Price of Anarchy of
% first-price auctions when the bidders have subadditive
% valuations~\cite{feldman_simultaneous_2012,CKST13}. In contrast,
% this similarity vanishes for the case of single item, as mixed
% equilibria in First Price auctions are efficient.

\end{abstract}
%%% Local Variables: 
%%% mode: latex
%%% TeX-master: "allpay"
%%% End: 

%% file: introduction.tex
\section{Introduction}

%We are interested in {\em all-pay auctions} where all players, even the losers, 
%pay their bid(s). 
It is a common economic phenomenon in competitions that agents make
irreversible investments without knowing the outcome. {\em
  All-pay} auctions are widely used in economics to
capture such situations, where all players, even the losers, pay their
bids. For example, a lobbyist can make a monetary contribution in
order to influence decisions made by the government. Usually the group
invested the most increases their winning chances, but all groups have
to pay regardless of the outcome. In addition, all-pay auctions have
been shown useful to model
% All-pay auctions are widely studied in economics and although they are
% not used in practice to sell ``real'' goods, they are very useful to model
rent seeking, political campaigns and R\&D races.
% lobbying activities \cite{BKV93}, rent seeking \cite{HS87}, 
% political campaigns \cite{Sny89}, job promotions \cite{Ros86}, 
% R\&D races \cite{CG03} and sporting contests \cite{Szy03}. 
%
There is a well-known connection between all-pay auctions and {\em
  contests}~\cite{Sie03}. % An {\em all-pay} contest \cite{Sie03} consists
% of selfish players competing for $m$ homogeneous prizes. Players
% simultaneously submit a score (single bid) and the $m$ highest scorers
% receive from one prize each. All players, even the losers, should pay
% an amount usually depending on their score.
In particular, the all-pay auction can be viewed as a single-prize
contest, where the payments correspond to the effort that players
make in order to win the competition.% ; for example, scientists invest
% research effort to secure patents.
% Obviously, these efforts are
% irreversible and all-pay auction can successgully model such kind of
% contests.

In this paper, we study the efficiency of mixed Nash equilibria in
all-pay auctions with complete information, from a worst-case analysis
perspective, using the {\em price of anarchy}~\cite{KP99} as a
measure. As social objective, we consider the {\em social welfare},
i.e. the sum of the bidders' valuations. We study the equilibria
induced from all-pay mechanisms in three fundamental resource
allocation scenarios; combinatorial auctions, multi-unit auctions and
single-item auctions.% (all-pay contests).

% The {\em social welfare}, i.e. the sum of the bidders' valuations, is
% a fundamental objective in economics. Our focus is on the analysis of
% the {\em equilibrium (in)efficiency} of all-pay mechanisms, i.e. an
% outcome is called {\em efficient} if its social welfare is maximized.
% Naturally, the equilibria can be inefficient, and we are interested in
%In all these mechanisms, all participants always pay their bids 
%regardless if they win the item(s) or not. 

In a combinatorial auction a set of items are allocated to a group of
selfish individuals. Each player has different preferences for
different subsets of the items and this is expressed via a {\em
  valuation set} function. % One common question is how to
% allocate the items among the participants so that the overall social
% efficiency is maximized. 
A multi-unit auction can be considered as an important special case,
where there are multiple copies of a single good. Hence the valuations
of the players are not set functions, but depend only on the number of
copies received. Multi-unit auctions have been extensively studied
since the seminal work by Vickrey~\cite{Vic61}. As already
mentioned, all-pay auctions have received a lot of attention for the
case of a single item, as they model all-pay contests and procurements
via contests.

\subsection{Contribution}

\noindent\emph{Combinatorial Auctions.} 
Our first result is on the price of anarchy of simultaneous all-pay
auctions with item-bidding that was previously studied by Syrgkanis
and Tardos \cite{ST13}. For 
fractionally subadditive valuations, it was previously shown that the
price of anarchy was at most $2$ \cite{ST13} and at least
$e/(e-1)\approx 1.58$ \cite{CKST13}.  % Our intention is to delve deeper
% in understanding of equilibria to improve the upper bounds on the
% Price of
% Anarchy.
% come up with an improved upper bound of $1.82$, for mixed Nash equilibria.
%and overcome the contingent limits of the upper bound approaches used in \cite{ST13,CKST13}. 
We narrow further this gap, by improving the upper bound to $1.82$. 
%we combine ideas from the upper bound techniques used in \cite{CKST13}, together with 
In order to obtain the bound, we come up with several structural
theorems that characterize mixed Nash equilibria in simultaneous
all-pay auctions.

\smallskip\noindent\emph{Multi-unit Auctions.} 
Our next result shows a novel use of all-pay mechanisms to the
multi-unit setting. We propose an all-pay mechanism with a randomized
allocation rule inspired by Kelly's seminal proportional-share
allocation mechanism~\cite{Kel97}. We show that this mechanism
admits a {\em unique}, $75\%$ efficient {\em pure} Nash equilibrium
and no other mixed Nash equilibria exist, when bidders' valuations are
submodular. As a consequence, the price of anarchy of our mechanism
outperforms all current price of anarchy bounds of prevalent
multi-unit auctions including uniform price auction~\cite{MT12} and
discriminatory auction~\cite{KMST13}, where the bound is
$e/(e-1)\approx 1.58$.

\smallskip\noindent\emph{Single-item Auctions.}   
Finally, we study the efficiency of a single-prize contest that can be
modeled as a single-item all-pay auction. We show a tight bound on the
price of anarchy for mixed equilibria which is approximately
$1.185$. By following previous study on the procurement via contest,
we further study two other standard
objectives, {\em revenue} and {\em maximum bid}. We evaluate the performance
of all-pay auctions in the prior-free setting, i.e. no distribution
over bidders' valuation is assumed. We show that both the revenue and
the maximum bid of any mixed Nash equilibrium are at least as high as $v_2/2$,
where $v_2$ is the second highest valuation. In contrast, the revenue and the maximum
bid in some mixed Nash equilibrium may be less than $v_2/2$ when using
reward structure other than allocating the entire reward to the
highest bidder. This result coincides with the optimal crowdsourcing
contest developed in \cite{CHS12} for the setting with prior
distributions. We also show that in conventional procurements (modeled
by first-price auctions), $v_2$ is exactly the revenue and maximum
bid in the worst equilibrium. So procurement via all-pay contests is a
$2$-approximation to the conventional procurement in the context of
worst-case equilibria.

%% file: relatedWork.tex
\subsection{Related work}
The inefficiency of Nash equilibria in auctions has been a well-known
fact (see e.g. \cite{Kri02}). Existence of efficient
equilibria of simultaneous sealed bid auctions in full information
settings was first studied by Bikhchandani~\cite{Bik99}.
% Koutsoupias and
% Papadimitriou \cite{KP99} initiated the study of quantifying the
% (in)efficiency of Nash Equilibrium which is also referred to be
% ``price of anarchy''\cite{Pap2001}. 
Christodoulou, Kov\'acs and Schapira~\cite{CKS08} initiated the study
of the (Bayesian) price of anarchy of simultaneous auctions with
item-bidding. %, where each item is assigned by a second-price auction. 
Several variants have
been studied since then~\cite{BR11,HKMN11,FFGL13}, as well as multi-unit auctions \cite{KMST13,MT12}. 
%and generalized second/first price auctions \cite{CKK11,PT10,CH13}). 

Syrgkanis and Tardos \cite{ST13} proposed
a general smoothness framework for several types of mechanisms and
applied it to settings with fractionally subadditive bidders obtaining
several upper bounds (e.g., first price auction, all-pay auction, 
and multi-unit auction). % and proportional allocation). 
%Feldman et
%al. \cite{FFGL13} showed tighter bounds for
%simultaneous first and second price auctions when the players have
%subadditive functions. 
%%by developing a technique different from smoothness. 
%The upper bounds for first-price auctions developed in the
%above two papers have been shown to be tight by Christodoulou
%et al.\cite{CKST13}. 
Christodoulou et al.~\cite{CKST13} constructed tight lower bounds for
first-price auctions and showed a tight price of anarchy bound of $2$ for all-pay
auctions with subadditive valuations. Roughgarden \cite{Rou14}
presented an elegant methodology to provide price of anarchy lower
bounds via a reduction from the hardness of % communication or computational complexity
% lower bounds of
the underlying optimization problems. % For auctions
% with subadditive and fractionally subadditive valuations, he provided
% general price of anarchy lower bounds of $2$ and $e/(e - 1)$,
% respectively, that apply also to item-bidding setting.
% as well as a lower bound for fractionally subadditive valuations.

%\subsubsection{All Pay Auction}
All-pay auctions and contests have been studied extensively in economic
theory.
%mainly for the case of a single item-prize, \cite{BKV96,PR10}, or as multi-prize auctions where each bidder may receive at most one item, \cite{BK07,CR98,Sie03}. There are also papers studying all-pay auctions with budget constraints, \cite{Kva07}. 
%In several works they compare all-pay auction with first price
%auction. Easley and Kleinberg, \cite{EK10}, study the Bayesian case
%with single item for both first price and all-pay auctions, where the
%values of the players are independently and uniformly distributed
%between $0$ and $1$. They derive an equilibrium for this case by
%assuming that all bidders follow the same bidding strategy, when they
%have the same value, and they further prove that their bidding is
%strictly increasing by the value. The analysis they use for the
%all-pay auction is very similar to the one for the first price
%auction. Krishna and Morgan, \cite{KM97}, compare the revenue of the
%all-pay and the first price auction when there is a single item and
%the bidders' valuations are affiliated and symmetrically
%distributed. They show that, under certain circumstances, the revenue
%of the all-pay auction always outperforms this of first price
%auction.  
Baye, Kovenock and de Vries \cite{BKV96}, fully characterized the Nash
equilibria in single-item all-pay auction with complete information.  
The connection between all-pay auctions and crowdsourcing contests was
proposed in \cite{DV09}. Chawla et al.~\cite{CHS12} studied the design of optimal
crowdsourcing contest to optimize the maximum bid in all-pay auctions
when agents' value are drawn from a specific distribution
independently. % The optimal design of crowdsourcing contests has been
% also considered in \cite{AS09}, when the objective is to maximize the
% highest $K$ bids.

%%% Local Variables: 
%%% mode: latex
%%% TeX-master: "allpay"
%%% End: 

%% file: preliminaries.tex
\section{Preliminaries}
\label{sec:preliminary}

%\paragraph{Combinatorial Auctions} 
In a {\em combinatorial auction}, $n$ {\em players} compete on $m$ {\em
  items}. Every player (or {\em bidder}) $i\in [n]$
has a valuation function $v_i: \{0,1\}^m \rightarrow \mathbb R^+$
which is monotone and normalized, that is, $\forall S \subseteq T
\subseteq [m]$, $v_i(S)\leq v_i(T),$ and $v_i(\emptyset)=0.$ The
outcome of the auction is represented by a tuple of $(\X,\p)$
where $\X=(X_1,\ldots, X_n)$ specifies the allocation of items ($X_i$
is the set of items allocated to player $i$) and $\p=(p_1,\ldots,p_n)$
specifies the buyers' payments ($p_i$ is the payment of player $i$ for
the allocation $\X$).
%In the \emph{Bayesian} setting, the valuation of
%each player $i$ is drawn from $V_i$, a set of possible valuations,
%according to some known distribution $D_i$. We assume that $D_i$
%are independent (but not necessarily identical) over the players. 
%In this work we consider the {\em full information, prior-free} setting, 
%where for each player $i$, $v_i$ is a specific valuation known by all other players; 
%however, the auction designer has no prior information about the valuations.
In the {\em simultaneous item-bidding} auction, every
player $i\in [n]$ submits a non-negative bid $b_{ij}$ for each item
$j\in [m].$ The items are then allocated by independent auctions, 
i.e. the allocation and payment rule for item $j$ only
depend on the players' bids on item $j$.
In a simultaneous {\em all-pay} auction the 
allocation
% \footnote{Our (upper bound) results hold for arbitrary randomized
% tie-breaking rules, as long as, for any fixed
% $\b=(b_1,\ldots,b_n),$ the probabilities for the players to get
% the item are fixed.}
and payment for each player is determined as follows: each item $j \in [m]$ is
allocated to the bidder $i^*$ with the highest bid for that item,
i.e. $i^* = \arg\max_{i}b_{ij}$, and each bidder $i$ is charged an
amount equal to $p_{i}=\sum_{j\in [m]}b_{ij}$. It is worth mentioning
that, for any bidder profile, there always exists a tie-breaking rule
such that mixed equilibria exist \cite{simon1990discontinuous}. 
% Our
% results in this paper does not depend on any tie-breaking rule but for 
% completeness, we specify a tie-breaking rule: the
% mechanism pick a single agent from all winners uniformly if there are
% multiple winners.
\begin{definition} [Valuations] Let $v: 2^{[m]}\rightarrow \mathbb R$ be a valuation function. 
Then $v$ is called
\begin{inparaenum}[\itshape a\upshape)]
\item {\em additive}, if $v(S)=\sum_{j\in S} v(\{j\});$
\item {\em submodular}% \footnote{Equivalently, submodular valuations are the valuations with
% {\em decreasing marginal values}, i.e. $v(\{j\}\cup T)-v(T)\leq
% v(\{j\}\cup S)-v(S)$ holds for any item $j$ and $S\subseteq T.$}
, if $v(S\cup T)+v(S\cap T)\leq
v(S)+v(T);$
\item {\em fractionally subadditive} or {\em XOS}, if $v$ is
determined by a finite set of additive valuations $\xi_k$
such that $v(S)= \max_{k} \xi_k(S)$.
\end{inparaenum}
\end{definition}
The classes of the above valuations are in increasing order of inclusion.

\smallskip\noindent\emph{Multi-unit Auction.} 		
In a multi-unit auction, $m$ copies of an item are sold to $n$
bidders. Here, bidder $i$ 's valuation is a function that depends on
the number of copies he gets. That is $v_i: \{0, 1,\dots, m\}
\rightarrow \mathbb R^+$ and it is non-decreasing and normalized, with
$v_i(0) = 0$. We say a valuation $v_i$ is {\em submodular}, if it has
non-increasing marginal values, i.e.  $v_i(s+1)-v_i(s)\geq
v_i(t+1)-v_i(t)$ for all $s\le t$.% such that $0\leq s\leq t \leq m.$

%The proportional-share allocation mechanism (PSAM) is used to 
%allocate one divisible resource to $n$ agents. 
%Given a bidding strategy profile, $\b=(b_1,\ldots,b_n)$, PSAM allocates  
%${b_{i}}/{\sum_{k\in[n]}b_{k}}$ amount of the resource to agent $i$, 
%and agent $i$ pay her own bid, $b_i$. 
		%When all agents bidding $0$, the allocation can be
    %defined arbitrarily but consistently.

\smallskip\noindent\emph{Nash equilibrium and price of anarchy.} We use $b_i$ to denote a pure
strategy of player $i$ which might be a single value or a vector,
depending on the auction. So, for the case of $m$ simultaneous
auctions, $b_i=(b_{i1},\ldots ,b_{im})$.
% For any allocation $\x_i$, we use the notation
% $b_i(x_i)=\sum_{j\in [m]}x_{ij}b_i(j)$ to denote the sum of the bids
% of player $i$ for allocation $x_i$.
We denote by $\b_{-i}=(b_1,\ldots,b_{i-1},b_{i+1},\ldots,b_n)$ the strategies
of all players except for $i.$ Any %\emph{mixed, correlated, coarse correlated or Bayesian strategy $B_i$}
{\em mixed strategy $B_i$}	of player $i$ is a
probability distribution over pure strategies.

For any profile of strategies,
$\b=(b_1,\ldots,b_n)$, $\X(\b)$ denotes the allocation under the
strategy profile $\b$. The valuation of player $i$ for the allocation 
$\X(\b)$ is denoted by $v_i(\X(\b))=v_i(\b)$. 
% respectively, since the allocation only depends on the auction's
% allocation rule (which is fixed before the auction takes place) and
% the strategy profile.
The {\em utility $u_i$} of player $i$ is defined as the difference
between her valuation %for the received allocation 
and payment:
$u_i(\X(\b))=u_i(\b)=v_i(\b)-p_i(\b)$.

\begin{definition} [Nash equilibria] A bidding profile $\b=(b_1,\ldots,b_n)$
  forms a pure Nash equilibrium if for every player $i$ 
	and all bids $b'_i$, $u_i(\b)\ge u_i(b'_i, \b_{-i})$. Similarly, a mixed
  bidding profile $\B=\times_i B_i$ is a mixed Nash equilibrium if 
	for all bids $b'_i$ and every player $i$, 
  $\E_{\b\sim\B}[u_i(\b)]\ge \E_{\b_{-i}\sim
    \B_{-i}}[u_i(b'_i,\b_{-i})]$.
  Clearly, any pure Nash equilibrium is also a mixed Nash equilibrium.
\end{definition}
% \emph{$\B=\left(B_i\right)_i$ and $\E_{\b\sim\B}[u_i(\b)|b_i]\ge
% \E_{\b\sim\B}[u_i(b'_i,\b_{-i})|b_i]$}&\emph{ (\emph{correlated
% equilibrium}),} 
% \emph{$\B=\left(B_i\right)_i$ and $\E_{\b\sim\B}[u_i(\b)]\ge
% \E_{\b\sim \B}[u_i(b'_i,\b_{-i})]$}&\emph{ (\emph{coarse correlated
% equilibrium}),} 
% \emph{$\B(\v)=\times_iB_i(v_i)$ and
% $\E_{\v_{-i},\b}[u_i(\b)]\ge \E_{\v_{-i},\b_{-i}}[u_i(b'_i,
% \b_{-i})]$}&\emph{ (\emph{Bayesian Nash equilibrium}).}
%\end{tabular}
% $\subseteq$ correlated $\subseteq$ coarse
%correlated and mixed Nash $\subseteq$ Bayesian Nash.

Our global objective is to maximize the sum of the valuations of the
players for their received allocations, i.e., to maximize the
{\em social welfare} $SW(\X)=\sum_{i\in [n]} v_i(X_i).$  
So $\O(\v)=\O=(O_1,\ldots,O_n)$ is an {\em optimal allocation} if $SW(\O)=\max_{\X} SW(\X)$. 
In Sect. \ref{sec:single}, we also study two other objectives: the {\em revenue}, which 
equals the sum of the payments, $\sum_i p_i$,
and the {\em maximum payment}, $\max_i b_i$. We also refer to the maximum payment 
as the {\em maximum bid}. 

%For simplicity, if the allocation rule $\X$ is clear from the context, 
%we use $SW(\b)$, $v_i(\b)$ and $u_i(\b)$ 
%instead of $SW(\X(\b))$, $v_i(X_i(\b))$ and $u_i(\X(\b))$, to express
%social welfare, valuation and utility of player $i$ for the allocation $\X(\b)$.

\begin{definition}[Price of anarchy]
Let $\mathcal{I}([n],[m],\v)$ be the set of all
instances, i.e. $\mathcal{I}([n],[m],\v)$ includes
the instances for every set of bidders and items and any possible
valuation functions. The mixed 
%, correlated, coarse correlated and Bayesian 
price of anarchy, PoA, of a mechanism is defined as
$$\mbox{PoA} = \max_{I \in \mathcal{I}} \max_{\B \in \mathcal{E}(I)}
\frac{SW(\O)}{\E_{\substack{
      \b\sim\B}}[SW(\X(\b))]}\enspace , $$ where $\mathcal{E}(I)$ is the class of
mixed Nash %, correlated, coarse correlated or Bayesian Nash
equilibria for the instance $I \in \mathcal{I}$. 
The pure PoA is defined as above but restricted in the
class of pure Nash equilibria.
% , 
% i.e. in the above formula, $\mathcal{E}(I)$ is the class of
% pure Nash equilibria. %  and for the definition of
% pure PoA there is no need of eny expectation.
\end{definition}

%For the case of single item, we similarly study the revenue and 
%maximum bid objectives, for strategy profiles that constitute mixed Nash equilibria. 
%We compare the outcomes for the first-price and all-pay auction, by 
%considering the second highest valuations as a benchmark. 

Let $\B=(B_1,\ldots,B_n)$ be a profile of mixed strategies. Given the
profile $\B$, we fix the notation for the following {\em cumulative
  distribution functions (CDF):} $G_{ij}$ is the CDF of the bid of
player $i$ for item $j;$ $F_{j}$ is the CDF of the highest bid for
item $j$ and $F_{ij}$ is the CDF of the highest bid for item $j$ if we
exclude the bid of player $i.$ Observe that $F_j=\prod_k G_{kj}$ and
$F_{ij}=\prod_{k\neq i} G_{kj}.$ We also use $\varphi_{ij}(x)$ to denote
the probability that player $i$ gets item $j$ by bidding $x.$ Then, 
$\varphi_{ij}(x)\leq F_{ij}(x).$ When we refer to a single item, we
may drop the index $j$. 
Whenever it is clear from the context, we will
use shorter notation for expectations, e.g. we use $\E[u_i(\b)]$
instead of $\E_{\b\sim \B}[u_i(\b)]$, or even 
%or for fixed $b'_i,$ we use $\E[u_i(b'_i)]$ instead of
%$\E_{\b_{-i}\sim \B_{-i}}[u_i(b'_i,\b_{-i})].$ 
$SW(\B)$ to denote $\E_{\b \sim \B}[SW(\X(\b))]$.

%Due to the
%space limitation, we put several proofs into Appendix \ref{ap:proofs}.

%%% Local Variables: 
%%% mode: latex
%%% TeX-master: "allpay"
%%% End: 

%% file: xos.tex
\section{Combinatorial Auctions}

\label{sec:all-pay-1.82}
In this section we prove an upper bound of $1.82$ for the mixed price of
anarchy of simultaneous all-pay auctions when bidders' valuations are
fractionally subadditive (XOS). This result improves over the previously
known bound of $2$ due to \cite{ST13}. We first state our main theorem
and present the key ingredients. Then we prove these ingredients in
the following subsections.

\begin{theorem}
  \label{thm:1.82} The mixed price of anarchy for simultaneous
  all-pay auctions with fractionally subadditive (XOS) bidders is at most $1.82$.
\end{theorem}

\begin{proof}
    Given a valuation profile $\v=(v_1,\ldots,v_n),$ let $\O =
  (O_1,\ldots , O_n)$ be a fixed optimal solution, that maximizes the
  social welfare. We can safely assume that $\O$ is a partition of the items. 
	Since $v_i$ is an XOS valuation,
  let $\xi^{O_i}_i$ be a maximizing additive function with respect to $O_i$. 
  For every item $j$ we denote by $o_j$
  item $j$'s contribution to the optimal social welfare, that is,
  $o_j=\xi^{O_i}_i(j)$, where $i$ is such that $j\in O_i$. The optimal social welfare is thus
  $SW(\O)=\sum_jo_j.$ In order to bound the price of anarchy, we consider only
  items with $o_j>0$, as it is without loss of
  generality to omit items with $o_j=0.$

  For a fixed mixed Nash equilibrium $\B,$ recall that by $F_j$ and $F_{ij}$
  we denote the CDFs of the maximum bid on item $j$ among all bidders,
  with and without the bid of bidder $i$, respectively.  %Observe that $F_j(x)\le F_{ij}(x)$. 
	For any item $j\in O_i$, let $A_j=\max_{x\geq 0} \,
  \{F_{ij}(x)o_j-x\}.$

  As a key part of the proof we use the following two inequalities
  that bound from below the social welfare in any mixed Nash
  equilibrium $\B$. 
\begin{eqnarray}
    SW(\B) &\ge& \sum_{j\in[m]}\left(A_j+\int_0^{o_j-A_j}(1-F_j(x))dx\right) \enspace , \label{ineq:1}\\
    SW(\B) &\ge& \sum_{j\in [m]}\int_0^{o_j-A_j}\sqrt{F_j(x)}dx \enspace .  \label{ineq:2}
\end{eqnarray}
  % \begin{lemma}
  %   \label{lem:inequality-one} 
  %   $SW(\B) \ge \sum_{j\in[m]}(A_j+\int_0^{o_j-A_j}(1-F_j(x))dx)$
  % \end{lemma}
  % \begin{lemma}
  %   \label{lem:inequality-two} 
  %   $SW(\B) \ge \sum_{j\in [m]}\int_0^{o_j-A_j}\sqrt{F_j(x)}dx$
  % \end{lemma}
Inequality \eqref{ineq:1} suffices to provide a weaker
upper bound of $2$ (see \cite{CKST13}). The proof of \eqref{ineq:2} is much
more involved, and requires a deeper understanding of the
equilibria properties of the induced game.  We postpone their proofs in
Sect.~\ref{sec:inequality1} (Lemma~\ref{lem:inequality-one}) and
Sect.~\ref{sec:inequality-two} (Lemma~\ref{lem:inequality-two}),
respectively.  
By combining \eqref{ineq:1} and \eqref{ineq:2},
\begin{equation}
\label{eq:aa}
SW(\B)\ge\frac{1}{1+\lambda}\cdot\sum_j\left(A_j+\int_0^{o_j-A_j}
\left(1-F_j(x)+\lambda \cdot \sqrt{F_j(x)}\right)dx\right) \enspace ,
\end{equation}
for every $\lambda\ge 0$. It suffices to bound from below the
right-hand side of \eqref{eq:aa} with respect to the optimal social
welfare. For any cumulative distribution function $F$, and any
positive real number $v$, let
$$R(F,v)\stackrel{\mathrm{def}}{=}A+\int_0^{v-A}(1-F(x))dx+\lambda\cdot\int_0^{v-A}\sqrt{F(x)}dx \enspace ,$$
where $A=\max_{x\geq 0}\{F(x)\cdot v - x\}$.
Inequality~\eqref{eq:aa} can then be rewritten as $SW(\B)\ge$ 
$\frac{1}{1+\lambda}\sum_jR(F_j,o_j)$. 
%Note that $F_j(x)\le
%F_{ij}(x)\le (A_j+x)/o_j$ and $A_j=\max_x \,\{F_{ij}(x)\cdot
%o_j-x\}\le o_j$. 
Finally, we show a lower bound of $R(F,v)$ that holds for any CDF $F$
and any positive real $v$.
%such that $F(x)\le (x+A)/v$. 

\begin{equation}
R(F,v)\ge \frac{3+4\lambda-\lambda^4}{6}\cdot v \enspace . \label{eq:CDF-ineq}
\end{equation}
The proof of \eqref{eq:CDF-ineq} is given in
Sect.~\ref{sec:inequality-optimal-cdf} (Lemma~\ref{lem:optimal-cdf}).
Finally, we obtain that for any $\lambda> 0$,
\[SW(\B)\ge \frac{1}{1+\lambda}\sum_jR(F_j,o_j)\ge
\frac{3+4\lambda-\lambda^4}{6\lambda+6}\cdot\sum_jo_j=
\frac{3+4\lambda-\lambda^4}{6\lambda+6}\cdot SW(\O) \enspace .\] 
By taking $\lambda=0.56$, we conclude that the price of anarchy is at most $1.82$. 
\end{proof}

\subsection{Proof of Inequality (\ref{ineq:1})}
\label{sec:inequality1}
This section is devoted to the proof of the following lower
bound. Recall that the definition $o_j$ is from the definition of XOS functions.
  \begin{lemma}
    \label{lem:inequality-one} 
    $SW(\B) \ge \sum_{j\in[m]}(A_j+\int_0^{o_j-A_j}(1-F_j(x))dx).$
  \end{lemma}

\begin{proof}
  Recall that $A_j=\max_{x_j\geq 0} \, \{F_{ij}(x)o_j-x_j\}$. We can bound
  bidder $i$'s utility in the Nash equilibrium $\B$ by $u_i(\B)\ge \sum_{j\in
    O_i}A_j$. To see this, consider the deviation for bidder $i$, where he bids 
		only for items in $O_i$, namely, for each item $j$, he bids the value $x_j$ 
		that maximizes the expression $F_{ij}(x_j)o_j-x_j$.
		%If this was not the case, bidder $i$, would bid only for
  %items in $O_i,$ namely for each item $j$ the bid $x$ that maximizes
  %$F_{ij}(x)o_j-x$, improving his utility, and contradicting the fact that
  %$\B$ is a Nash equilibrium.  
	Since for any obtained subset $T\subseteq
  O_i$, he has value $v_i(T)\geq \sum_{j\in T} o_j,$ and the bids $x_j$ must
  be paid in any case, the expected utility with these bids is at
  least $\sum_{j\in O_i}\max_{x_j\geq 0} \,(F_{ij}(x)o_j-x_j)=\sum_{j\in
    O_i}A_j.$ With $\B$ being an equilibrium, we infer that $u_i(\B)\ge \sum_{j\in
    O_i}A_j$.
  %Here, we also used that $\varphi_{ij}=F_{ij}$ by Lemma~\ref{lem:no-ties}.
%\end{proof}
% \begin{claim}\label{lem:sumA} 
%   For every bidder $i$, $u_i(\B)\ge \sum_{j\in O_i}A_j$.
% \end{claim}
% \begin{proof} 
%
% By above lemma, we can show a lower bound for the social welfare in
% $\B$ by using the fact that utility is the difference between
% valuation and payment.
%
%\begin{proof}[of Lemma \ref{lem:inequality-one}]
  By summing up over all bidders, 
  \begin{align*} 
    SW(\B)=&\sum_{i\in[n]}u_i(\B)+\sum_{i\in[n]}\sum_{j\in[m]}\E[b_{ij}]
    \ge\sum_{j\in[m]}A_j+\sum_{j\in[m]}\sum_{i\in[n]}\E[b_{ij}]\\ 
    \ge&\sum_{j\in[m]}(A_j+\E[\max_{i\in[n]}\{b_{ij}\}])
    \ge\sum_{j\in[m]}\left(A_j+\int_0^{o_j-A_j}(1-F_j(x))dx\right) \enspace .
  \end{align*}
  The first equality holds because 
  $\sum_i\E_\b[v_i(\b)]=\sum_{i}\E_\b[u_i(\b)+\sum_{j\in[m]}b_{ij}].$
  The second inequality follows because 
  $\sum_{i} b_{ij}\geq \max_{i} \,b_{ij}$ and the last one is
  implied by the definition of the expected value of any
  positive random variable. %$Y$ with CDF given by $F.$
%?? is the last sentence correct??
\end{proof}

\subsection{Proof of Inequality (\ref{ineq:2})}
\label{sec:inequality-two}
Here, we prove the following lemma for any mixed Nash equilibrium $\B$.
  \begin{lemma}
    \label{lem:inequality-two} 
    $SW(\B) \ge \sum_{j\in [m]}\int_0^{o_j-A_j}\sqrt{F_j(x)}dx.$
  \end{lemma}
	
First we show a useful lemma that holds for XOS valuations. 
We will further use the technical Proposition \ref{prop:inequality}, whose 
proof is deferred to Appendix~\ref{sec:inequality}.

\begin{lemma}
  \label{lem:xos}
  For any fractionally subadditive (XOS) valuation function $v$,
  $$v(S) \ge \sum_{j\in [m]}\left(v(S)-v(S\setminus\{j\})\right) \enspace .$$
\end{lemma}
\begin{proof}
  Let $\xi$ be a maximizing additive function of $S$ for the XOS valuation
  $v$. By definition, $v(S)=\xi(S)$ and for every $j$, 
  $v(S\setminus\{j\})\ge \xi(S\setminus\{j\})$. Then,
  $\sum_{j\in [m]}\left(v(S)-v(S\setminus\{j\})\right)\le \sum_{j\in
    S}(\xi(S)-\xi(S\setminus\{j\}))=\sum_{j\in
    S}\xi(j)=v(S).$
  %The last equality is due to the additivity of $\xi$.
\end{proof}

\begin{proposition}
  \label{prop:inequality}
  For any integer $n\ge 2$, any positive reals $G_i\le 1$ and positive
  reals $g_i$, for $1\leq i\leq n,$
  $$\sum_{i=1}^n\frac{g_i}{\sum_{k\neq i}\frac{g_k}{G_k}}
  \ge\sqrt{\prod_{i=1}^nG_i} \enspace .$$
\end{proposition}

We are now ready to prove 
Lemma~\ref{lem:inequality-two}. We first state a proof sketch here
to illustrate the main ideas.

%?? I changed the proof
\begin{proof} [Sketch of Lemma~\ref{lem:inequality-two}]
%?? can you give a reference to literatue here??
  Recall that $G_{ij}$ is the CDF of the bid of player $i$ for item
  $j$. For simplicity, we assume $G_{ij}(x)$ is continuous and
  differentiable, with $g_{ij}(x)$ being the PDF of player $i$'s bid
  for item $j$. The general case will be considered later. First,
  we define the {\em expected marginal valuation} of item $j$ w.r.t
  player $i$,
  $$v_{ij}(x)\stackrel{\mathrm{def}}{=}\E_{\b\sim\B}[v_i(X_i(\b)\cup\{j\})-v_i(X_i(\b)\setminus\{j\})|b_{ij}=x] \enspace .$$
  Given the above definition and a careful characterization of mixed Nash
  equilibria, we are able to show
  $F_{ij}(x)\cdot
  v_{ij}(x)=\E[v_i(X_i(\b))-v_i(X_i(\b)\setminus\{j\})|b_{ij}=x]$
  and $\frac{1}{v_{ij}(x)}=\frac{dF_{ij}(x)}{dx}$ for any $x$ in the
  support of $G_{ij}$. Let $g_{ij}(x)$ be the derivative of
  $G_{ij}(x)$. Using Lemma \ref{lem:xos}, we have
  \begin{align*} 
    SW(\B)&=\sum_i\E[v_i(X_i(\b))]
            \ge\sum_i\sum_j\E[v_i(X_i(\b))-v_i(X_i(\b)\setminus\{j\})]\\
    &\ge\sum_i\sum_j\int_{0}^{o_j-A_j}\E[v_i(X_i(\b))-v_i(X_i(\b)\setminus\{j\})|b_{ij}=x]\cdot
    g_{ij}(x)dx\\ 
    &\ge\sum_i\sum_j\int_0^{o_j-A_j}F_{ij}(x)\cdot v_{ij}(x)\cdot g_{ij}(x)dx \enspace ,
  \end{align*}
  where the second inequality follows by the law of total probability. By
  using the facts that $F_{ij}(x)=\prod_{k\neq i}G_{kj}(x)$ and 
  $\frac{1}{v_{ij}(x)}=\frac{dF_{ij}(x)}{dx}$, for any $x>0$ such that $g_{ij}(x)>0$ ($x$ is in the support of player $i$) and 
  $F_{j}(x)>0$, we obtain
  \begin{align*} 
    F_{ij}(x)\!\cdot \! v_{ij}(x)\! \cdot \! g_{ij}(x)\!=&\frac{F_{ij}(x)\!\cdot\! g_{ij}(x)}{\frac{dF_{ij}}{dx}(x)}
    \!=\!\frac{\prod_{k\neq i}G_{kj}(x)\!\cdot\! g_{ij}(x)}{\sum_{k\neq i}\!\left(g_{kj}\!\cdot\!\prod_{s\neq k \wedge s\neq i}G_{sj}\right)}
    \!=\!\frac{g_{ij}(x)}{\sum_{k\neq i}\frac{g_{kj}(x)}{G_{kj}(x)}} \enspace \! .
  \end{align*} 
  For every $x>0$, we use Proposition~\ref{prop:inequality} only over the set $S$ of players with $g_{ij}(x)>0$.  
	%(in Appendix \ref{app:inequality-two} we argue that $|S| \geq 2$)
	After summing over all bidders we get,
  $$\sum_{i\in[n]} F_{ij}(x)\cdot v_{ij}(x)\cdot g_{ij}(x)\ge\sum_{i\in
    S}\frac{g_{ij}(x)}{\sum_{k\neq
      i,k\in S}\frac{g_{kj}}{G_{kj}}}\ge\sqrt{\prod_{i\in S}G_{ij}(x)} \ge \sqrt{F_j(x)} \enspace .$$
		The above inequality also holds for $F_j(x)=0$. 
Finally, by merging the above inequalities, we conclude that 
  $SW(\B)\ge \sum_{j\in[m]}\int_0^{o_j-A_j}\sqrt{F_j(x)}dx.$ 
\end{proof}

Now we show the complete proof for Lemma~\ref{lem:inequality-two}.
Recall that $o_j$ is the contribution of item $j$ to the optimum
social welfare. If player $i$ is the one receiving item $j$ in the
optimum allocation, then
$A_j = \max_{x\geq 0} \{F_{ij}(x)\cdot o_j-x\}$. The proof of
Lemma~\ref{lem:inequality-two} needs a careful technical preparation
that we divided into a couple of lemmas.

First of all, we define the expected marginal
valuation of item $j$ for player $i.$ For given mixed strategy
$B_i,$ the distribution of bids on items in $[m]\setminus \{j\}$
depends on the bid $b_{ij},$ so one can consider the  given
conditional expectation: 

\begin{definition}
  Given a mixed bidding profile $\B=(B_1, B_2, \ldots, B_n)$, the
  expected marginal valuation $v_{ij}(x)$ of item $j$ for player
  $i$ when $b_{ij}=x$ is defined as 
  $$v_{ij}(x)\stackrel{\mathrm{def}}{=}\E_{\b\sim\B}[v_i(X_i(\b)\cup\{j\})-v_i(X_i(\b)\setminus\{j\})|b_{ij}=x] \enspace .$$
\end{definition}
For a given $\B$, let $\varphi_{ij}(x)$ denote the probability that
bidder $i$ gets item $j$ when she bids $x$ on item $j$. It is
clear that $\varphi_{ij}$ is non-decreasing and $\varphi_{ij}(x)\le
F_{ij}(x)$ (they are equal when no ties occur).

\begin{lemma} \label{lem:phiv} For a given $\B$, for any bidder $i$, item $j$ and bids
  $x\geq 0$ and $y\geq 0$,
  \[\varphi_{ij}(y)\cdot
  v_{ij}(x)=\E_{\b\sim\B}[v_i(X_i(\b'))-v_i(X_i(\b')\setminus\{j\})|b_{ij}=x] \enspace ,\]
  where $\b'$ is the modified bid of $\b$ such that $\b'=\b$ except that
  $b'_{ij}=y$.
\end{lemma}
\begin{proof}
  \begin{eqnarray*} 
    &&\E_{\b\sim\B}[v_i(X_i(\b'))-v_i(X_i(\b')\setminus\{j\})|b_{ij}=x]\\
    &=&
    \E_{\b\sim\B}[v_i(X_i(\b'))-v_i(X_i(\b')\setminus\{j\})|b_{ij}=x, j
    \in X_i(\b')]\pr(j \in X_i(\b')|b_{ij}=x)\\
    &+& \E_{\b\sim\B}
    [v_i(X_i(\b'))-v_i(X_i(\b')\setminus\{j\})|b_{ij}=x, j \notin
    X_i(\b')]\pr(j \notin X_i(\b')|b_{ij}=x)\\
    &=&
    \E_{\b\sim\B}[v_i(X_i(\b'))-v_i(X_i(\b')\setminus\{j\})|b_{ij}=x, j
    \in X_i(\b')]\pr(j \in X_i(\b')|b_{ij}=x)\\
    &=& \E_{\b\sim\B}
    [v_i(X_i(\b'))-v_i(X_i(\b')\setminus\{j\})|b_{ij}=x, j \in
    X_i(\b')]\cdot \varphi_{ij}(y)\\
    &=& \E_{\b\sim\B}
    [v_i(X_i(\b')\cup \{j\})-v_i(X_i(\b')\setminus\{j\})|b_{ij}=x, j
    \in X_i(\b')]\cdot \varphi_{ij}(y)\\
    &=& \E_{\b\sim\B}
    [v_i(X_i(\b')\cup \{j\})-v_i(X_i(\b')\setminus\{j\})|b_{ij}=x]\cdot \varphi_{ij}(y)\\
    &=& \varphi_{ij}(y)\cdot v_{ij}(x) \enspace .
\end{eqnarray*}
The second equality is due to
$\E_{\b\sim\B}[v_i(X_i(\b')))-v_i(X_i(\b')\setminus\{j\})|b_{ij}=x,
j\notin X_i(\b')]=0;$ the third one holds because $b'_{ij}=y,$ and that other players' bids have distribution $\times_{k\neq i} B_k.$ The fourth one is obvious, since $X_i(\b')=X_i(\b')\cup \{j\}$ given that $j\in X_i(\b').$  The last two equalities follow from the fact
that $v_i(X_i(\b')\cup\{j\})-v_i(X_i(\b')\setminus\{j\})$ is
independent of the condition $j\in X_i(\b')$ and of the player $i$'s bid on
item $j$. 
\end{proof}

\begin{definition}
  Given a Nash equilibrium $\B$, we say a bid $x$ is {\em good for bidder
  $i$ and item $j$} (or $b_{ij}=x$ is {\em good}) if $\E[u_i(\b)]=\E[u_i(\b)|b_{ij}=x]$, otherwise
  we say $b_{ij}=x$ is {\em bad}.
\end{definition}
\begin{lemma}
  \label{lem:no-bad} Given a Nash equilibrium $\B$, for any bidder $i$
  and any item $j$, 
	$\pr[b_{ij}\mbox{ is bad }]=0$.
\end{lemma}
\begin{proof}
  The lemma follows from the definition of Nash equilibrium; otherwise we
  can replace the bad bids with good bids and improve the bidder's
  utility. 
\end{proof}

%?? added phi>0 to the lemma

\begin{lemma}
  \label{lem:opt-single}
  Given a Nash equilibrium $\B$, for any bidder $i$, item $j$, good
  bid $x$ and any bid $y\ge 0$,
  $$\varphi_{ij}(x)\cdot v_{ij}(x)-x \ge \varphi_{ij}(y)\cdot v_{ij}(x)-y \enspace .$$
Moreover, for a good bid $x>0,$ $\varphi_{ij}(x)>0$ holds.
\end{lemma}
\begin{proof}
  Let $\b'$ be the modified bid of $\b$ such that $\b'=\b$ except that
  $b'_{ij}=y$.
  \[\E[u_i(\b)]= \E[u_i(\b)|b_{ij}=x]\ge\E[u_i(\b')|b_{ij}=x] \enspace .\]
  Now we consider the difference between the above two terms:
  \begin{eqnarray*}
    0&\le&\E[u_i(\b)|b_{ij}=x]-\E[u_i(\b')|b_{ij}=x]\\
    &=&\E[v_i(X_i(\b))-b_{ij}|b_{ij}=x]-\E[v_i(X_i(\b'))-b'_{ij}|b_{ij}=x]\\ 
    &=&\E[v_i(X_i(\b))-v_i(X_i(\b)\setminus\{j\})|b_{ij}=x]\\
		&& -\E[v_i(X_i(\b'))-v_i(X_i(\b')\setminus\{j\}|b_{ij}=x]+y-x\\
    &=&(\varphi_{ij}(x)\cdot v_{ij}(x)-x)-(\varphi_{ij}(y)\cdot v_{ij}(x)-y) \enspace .
  \end{eqnarray*}
The second equality holds since $X_i(\b)\setminus\{j\}=X_i(\b')\setminus\{j\};$ the third equality holds by Lemma~\ref{lem:phiv}.

Finally, $\varphi_{ij}(x)>0$ for positive good bids follows by taking $y=0,$ 
since with $\varphi_{ij}(x)=0$ the left hand side of the inequality would be negative. 
\end{proof}
Next, by using the above lemma, we are able to show several structural
results for Nash equilibria. 

\begin{definition}
  \label{def:support}
  Given a mixed strategy profile $\B$, we say that a positive bid $x>0$ is
  in bidder $i$'s {\em support on item $j$}, if for all 
  $\varepsilon>0$, $G_{ij}(x)-G_{ij}(x-\varepsilon)>0$.
\end{definition}

\begin{lemma}
  \label{lem:good-bid}
  Given a mixed strategy profile $\B$, if a positive bid $x$ is in
  bidder $i$'s support on item $j$, then for every $\varepsilon>0$,
  there exists $x-\varepsilon< x' \le x$ such that $x'$ is good.
\end{lemma}
\begin{proof}
  Suppose on the contrary that there is an $\varepsilon>0$ such that for all $x'$, such that 
  $x-\varepsilon< x' \le x$, $x'$ is bad. Then $\pr[b_{ij}\mbox{
    is bad}]\ge G_{ij}(x)-G_{ij}(x-\varepsilon)>0$ (given that $x$ is in the support), which contradicts
  Lemma \ref{lem:no-bad}. 
\end{proof}

\begin{lemma}
  \label{lem:another-bidder}
  Given a Nash equilibrium $\B$, if $x>0$ is in bidder $i$'s support on
  item $j$, then there must exist another bidder $k\neq i$ such that
  $x$ is also in the bidder $k$'s support on item $j$, i.e. for all
  $\varepsilon>0$,
  $G_{kj}(x)-G_{kj}(x-\varepsilon)>0$.
\end{lemma}
\begin{proof}
  Assume on the contrary that for each player $k \neq i$, there exists
  $\varepsilon_k>0$ such that $G_{kj}(x)-G_{kj}(x-\varepsilon_k)=0$. 
	Clearly, for  $\varepsilon=\min\{\varepsilon_k|k\neq i\}$ 
	it holds that $G_{kj}(x)-G_{kj}(x-\varepsilon)=0$ for all bidders $k\neq i.$ That is
  $\varphi_{ij}(x)=\varphi_{ij}(x-\varepsilon)$. By Lemma
  \ref{lem:good-bid}, there exists $x-\varepsilon< x' \le x$ such that
  $x'$ is good for player $i$. Since $\varphi_{ij}$ is a non-decreasing function and
  $\varphi_{ij}(x)=\varphi_{ij}(x-\varepsilon)$, we have
  $\varphi_{ij}(x')=\varphi_{ij}(x-\varepsilon)$. By Lemma
  \ref{lem:opt-single}, $\varphi_{ij}(x')\cdot v_{ij}(x')-x' \ge
  \varphi_{ij}(x-\varepsilon)\cdot v_{ij}(x')-x+\varepsilon$ which
  contradicts the fact that
  $\varphi_{ij}(x')=\varphi_{ij}(x-\varepsilon)$ and
  $x'>x-\varepsilon$. 
\end{proof}

\begin{lemma}
  \label{lem:no-mass}
  Given a Nash equilibrium $\B$, for bidder $i$ and item $j$, there
  are no $x>0$ such that $\pr[b_{ij}=x]>0$, i.e. there are no mass
  points in the bidding strategy, except for possibly $0.$
\end{lemma}
\begin{proof}
  Assume on the contrary that there exists a bid $x>0$ such that
  $\pr[b_{ij}=x]>0$ for some bidder $i$ and item $j$. By Lemma
  \ref{lem:no-bad}, $x$ is good for bidder $i$ and item $j$, and  $\varphi_{ij}(x)>0$ by Lemma \ref{lem:opt-single}.

  According to Lemma \ref{lem:another-bidder}, there must exist a bidder $k$
  such that $x$ is in her support on item $j$. 
We can pick a sufficiently small $\varepsilon$ such that
  $\varepsilon<(x-\varepsilon)\cdot\varphi_{ij}(x)\cdot
  \pr[b_{ij}=x].$ This can be done since $(x-\varepsilon)$ 
   increases when $\varepsilon$ decreases.
Due to Lemma~\ref{lem:good-bid} there exists
  $x-\varepsilon< x' \le x$ such that $x'$ is good for bidder $k$ and
  item $j$. 
%By Lemma~\ref{lem:good-bid} for every $\varepsilon$ there exists
 % $x-\varepsilon< x' \le x$ such that $x'$ is good for bidder $k$ and
  %item $j$. If $x$ is good for $k,$ then let $x'=x,$ and let $\varepsilon<x\cdot\varphi_{ij}(x)\cdot
  %\pr[b_{ij}=x].$ Otherwise
 %we can pick a sufficiently small $\varepsilon$ such that
  %$\varepsilon<(x-\varepsilon)\cdot\varphi_{ij}(x-\varepsilon)\cdot
  %\pr[b_{ij}=x]$. This can be done since $(x-\varepsilon)$ and
  %$\varphi_{ij}(x-\varepsilon)$ increase when $\varepsilon$ decreases. (Note that $\varphi_{ij}(x-\varepsilon)>0$ for small enough $\varepsilon,$ given that an $x'<x$ with $\varphi_{kj}(x')>0$ exists, arbitrarily close to $x$). In this case fix some $x'>x-\varepsilon$ that is good for player $k.$
Now we consider the following two cases for $x'.$

  Case 1: $v_{kj}(x')\le x'$. Then $\varphi_{kj}(x')\cdot
  v_{kj}(x')-x'\le \varphi_{kj}(x')\cdot
  x'-x'\le(1-\varphi_{ij}(x)\cdot \pr[b_{ij}=x])\cdot x' - x'<0,$
  contradicting Lemma~\ref{lem:opt-single}. The first inequality holds
  by the case assumption. The second holds because player $k$ cannot
  get item $j$ with bid $x'$ whenever player $i$ gets it by bidding
  $x.$ The last inequality holds because both $\varphi_{ij}(x)>0$ and
  $\pr[b_{ij}=x]>0.$

  Case 2: $v_{kj}(x')> x'$. Then there exists a sufficiently small
  $\varepsilon'$ such that $\varepsilon'\le
  (x-\varepsilon)\cdot\varphi_{ij}(x)\cdot
  \pr[b_{ij}=x]-\varepsilon$. So $\varepsilon+\varepsilon'\le
  x'\cdot\varphi_{ij}(x)\cdot \pr[b_{ij}=x]$. Then,
  \begin{align*}
    &\varphi_{kj}(x+\varepsilon')\cdot v_{kj}(x')-x-\varepsilon'\\
    \ge& (\varphi_{kj}(x')+\varphi_{ij}(x)\cdot\pr[b_{ij}=x])\cdot
    v_{kj}(x')-x-\varepsilon'\\
    >& \varphi_{kj}(x')\cdot v_{kj}(x')+\varphi_{ij}(x)\cdot\pr[b_{ij}=x]\cdot x'-x'-(x-x')-\varepsilon'\\
    >& \varphi_{kj}(x')\cdot v_{kj}(x')+\varphi_{ij}(x)\cdot\pr[b_{ij}=x]\cdot x'-x'-\varepsilon-\varepsilon'\\
    \ge& \varphi_{kj}(x')\cdot v_{kj}(x')-x' \enspace ,
  \end{align*}
which contradicts Lemma \ref{lem:opt-single}. Here the first
inequality holds because the probability that player $k$ gets the item
with bid $x+\varepsilon'$ is at least  the probablity that he gets it
by bidding $x'$ plus the probability that $i$ bids $x$ and gets the
item (these two events for $\b_{-k}$ are disjoint). The second
inequality holds by case assumption, and the rest hold by our
assumptions on $\varepsilon$ and $\varepsilon'.$ 
\end{proof}

\begin{lemma}
  \label{lem:no-ties}
  Given a Nash equilibrium $\B$, for any bidder $i$ and item $j$,
  $\varphi_{ij}(x)=F_{ij}(x)$ for all $x>0$.
\end{lemma}
\begin{proof}
  The lemma follows immediately from Lemma \ref{lem:no-mass}. The
  probablity that some player $k\neq i$ bids exactly $x$ is zero. Thus
  $F_{ij}(x)$ equals the probability that the highest bid of players
  other than $i$ is strictly smaller than $x,$ and $1-F_{ij}(x)$ is
  the probability that it is strictly higher. Therefore $\varphi_{ij}(x)=F_{ij}(x).$ 
\end{proof}

\begin{lemma}
  \label{lem:value-mono}
  Given a Nash equilibrium $\B$, for any bidder $i$, item $j$ and good
  bids $x_1>x_2>0$, $v_{ij}(x_1)\ge v_{ij}(x_2)$.
\end{lemma}
\begin{proof}
  By Lemma \ref{lem:opt-single}, we have
  $(\varphi_{ij}(x_1)-\varphi_{ij}(x_2))\cdot v_{ij}(x_1)\ge x_1-x_2$
  and $(\varphi_{ij}(x_2)-\varphi_{ij}(x_1))\cdot v_{ij}(x_2)\ge
  x_2-x_1$. Combining these two inequalities, we have
  $$\frac{1}{v_{ij}(x_1)}\le
  \frac{\varphi_{ij}(x_1)-\varphi_{ij}(x_2)}{x_1-x_2}\le \frac{1}{v_{ij}(x_2)} \enspace .$$
\end{proof}

\begin{lemma}
  \label{lem:support-mono}
  Given a Nash equilibrium $\B$ and item $j$, let $T=\sup\{x|x$ is in
  some bidder's support on item $j\}$. For any bid $x<T$, $x$ is in some
  bidder's support on item $j$.
\end{lemma}
\begin{proof}
  Assume on the contrary that there exist a bid $x<T$ such that
  $x$ is not in any bidder's support. Then there exists $\delta>0$
  such that $G_{ij}(x)=G_{ij}(x-\delta)$ for all bidder $i$. Let
  $y=\sup\{z|\forall i, G_{ij}(x)=G_{ij}(z)\}$. By Lemma
  \ref{lem:no-mass}, $G_{ij}$ is continuous. So we have
  $G_{ij}(y)=G_{ij}(x)=G_{ij}(x-\delta)$ for any bidder $i$. That is 
  $F_{ij}(y)=F_{ij}(x-\delta)$ for any bidder $i$.

  By the definition of supremum, there exits a bidder $k$ such that
  for any $\varepsilon>0$,
  $G_{kj}(y+\varepsilon)>G_{kj}(x)=G_{kj}(y)$. By Lemma
  \ref{lem:no-bad}, there exists a good bid $y^+\in (y,y+\varepsilon]$
  for bidder $k$ and item $j$. We pick a sufficient small
  $\varepsilon$ such that $(F_{kj}(y^+)-F_{kj}(y))\cdot
  v_{kj}(y^+)<\delta$. This can be done since $F_{kj}$ is continuous
  by Lemma \ref{lem:no-mass} and $v_{kj}$ is non-decreasing by Lemma \ref{lem:value-mono}.
  \begin{align*}
    &F_{kj}(x-\delta)\cdot v_{ij}(y^+)-x+\delta\\
     =&F_{ij}(y)\cdot v_{ij}(y^+)-x+\delta\\
     >&F_{ij}(y)\cdot v_{ij}(y^+)-y^++\delta\\
     >&F_{ij}(y^+)\cdot v_{ij}(y^+)-y^+ \enspace ,
  \end{align*}
  which contradicts  Lemma \ref{lem:opt-single} and Lemma
  \ref{lem:no-ties}. 
  % there must exist a bid $y-\varepsilon<y'\le y$ such that $y'$ is
  % not in any bidder's support, i.e. there exists $\delta>0$ such
  % that $G_{kj}(y')=G_{kj}(y'-\delta)$ for any bidder $k$. So we have
  % $F_{ij}(y'-\delta)=F_{ij}(y')$. By Lemma \ref{lem:no-mass},
  % $F_{ij}$ is continuous for all positive points. So by taking
  % sufficient small $\varepsilon$, we have $F_{ij}(y)\cdot v_{ij}(y)<
  % F_{ij}(y-\varepsilon)\cdot v_{ij}(y)+\delta$. Then
  % \begin{align*}
  %   &F_{ij}(y'-\delta)\cdot v_{ij}(y)-y'+\delta\\
  %   =&F_{ij}(y')\cdot v_{ij}(y)-y'+\delta\\
  %   \ge&F_{ij}(y-\varepsilon)\cdot v_{ij}(y)-y+\delta\\
  %   >&F_{ij}(y)\cdot v_{ij}(y)-y
  % \end{align*}
  % which contradicts with Lemma \ref{lem:opt-single} and Lemma
  % \ref{lem:no-ties}. The first inequality is due to that fact that
  % $y-\varepsilon<y'\le y$.
\end{proof}

\begin{lemma}
  \label{lem:margin-derivative}
  Given a Nash equilibrium $\B$, if $x>0$ is a good bid for  bidder $i$ and item $j$, and
  $F_{ij}$ is differentiable in $x,$ then $$\frac{1}{v_{ij}(x)}=\frac{dF_{ij}(x)}{dx} \enspace .$$
\end{lemma}
\begin{proof} Notice that $v_{ij}(x)\neq 0$ by Lemma~\ref{lem:opt-single}.
  By Lemma \ref{lem:opt-single} and \ref{lem:no-ties}, we have
  $F_{ij}(x)\cdot v_{ij}(x)-x \ge F_{ij}(y)\cdot
  v_{ij}(x)-y$ for all $y\ge 0$. So for any $\varepsilon>0$,
  $$F_{ij}(x)\cdot v_{ij}(x)-x \ge F_{ij}(x+\varepsilon)\cdot
  v_{ij}(x)-x-\varepsilon \enspace ,$$
  $$F_{ij}(x)\cdot v_{ij}(x)-x \ge F_{ij}(x-\varepsilon)\cdot
  v_{ij}(x)-x+\varepsilon \enspace .$$
  That is,
  $$\frac{F_{ij}(x+\varepsilon)-F_{ij}(x)}{\varepsilon}\le \frac{1}{v_{ij}(x)} \enspace ,$$
  $$\frac{F_{ij}(x)-F_{ij}(x-\varepsilon)}{\varepsilon}\ge
  \frac{1}{v_{ij}(x)} \enspace .$$
  The lemma follows by taking the limit when $\varepsilon$ goes to $0$. 
\end{proof}

\begin{proof} [of Lemma~\ref{lem:inequality-two}]
%?? can you give a reference to literatue here??
  Since $G_{ij}(x)$ is non-decreasing, continuous (Lemma
  \ref{lem:no-mass}) and bounded by $1$, $G_{ij}(x)$
  is differentiable on almost all points. That is, the set of all
  non-differentiable points has Lebesgue measure $0$. So it will
  not change the value of integration if we remove these
  points. Therefore it is without loss of generality to assume
  $G_{ij}(x)$ is differentiable for all $x$. Let $g_{ij}(x)$ be the
  derivative of $G_{ij}(x)$, i.e. probability density function for
  bidder $i$'s bidding on item $j$. Using Lemma \ref{lem:xos}, we have
  \begin{align*} 
    SW(\B)&=\sum_i\E[v_i(X_i(\b))]\\ 
    &\ge\sum_i\sum_j\E[v_i(X_i(\b))-v_i(X_i(\b)\setminus\{j\})]\\
    &\ge\sum_i\sum_j\int_{0}^{o_j-A_j}\E[v_i(X_i(\b))-v_i(X_i(\b)\setminus\{j\})|b_{ij}=x]\cdot
    g_{ij}(x)dx\\ 
    &\ge\sum_i\sum_j\int_0^{o_j-A_j}F_{ij}(x)\cdot v_{ij}(x)\cdot g_{ij}(x)dx \enspace . 
  \end{align*}

  The second inequality follows by the law of total probability, and
  the third is due to Lemmas~\ref{lem:phiv} and \ref{lem:no-ties}.  By
  Lemma \ref{lem:margin-derivative} and the fact that
  $F_{ij}(x)=\prod_{k\neq i}G_{kj}(x)$, if $x$ is good, $g_{ij}(x)>0$
  and $G_{ij}(x)>0$ we have for all $j$ 
  \begin{align*} 
    F_{ij}(x)\cdot v_{ij}(x)\cdot g_{ij}(x)=&\frac{F_{ij}(x)\cdot g_{ij}(x)}{\frac{dF_{ij}}{dx}(x)}\\
    =\frac{\prod_{k\neq i}G_{kj}(x)\cdot g_{ij}(x)}{\sum_{k\neq i}\left(g_{kj}\cdot\prod_{s\neq k \wedge s\neq i}G_{sj}\right)}
    =&\frac{g_{ij}(x)}{\sum_{k\neq i}\frac{g_{kj}(x)}{G_{kj}(x)}} \enspace .
  \end{align*} 

	By concentrating on a specific item $j$, let $S_x$ be the set of bidders so that $x$
  is in their support. We next show that $|S_x| \geq 2$ for all $x\in (0, o_j-A_j]$.
	Recall that $A_j=\max_x \,\{F_{ij}(x)\cdot o_j-x\}$ for the bidder
  $i$ who receives $j$ in $\O.$ Let $h_{ij}=\min\{ x | F_{ij}=1\}$ 
	(we use minimum instead of infimum, since, by Lemma
  \ref{lem:no-mass}, $F_{ij}$ is continuous). By definition $h_{ij}$ 
	should be in some bidder's support. 
	Moreover, $A_j \geq F_{ij}(h_{ij})\cdot o_j-h_{ij} = o_j-h_{ij}$, resulting 
	in $o_j - A_j \leq h_{ij}$. 
	Therefore, by Lemma \ref{lem:support-mono}, for all
  $x\in (0, o_j-A_j]$, $\,x$ is in some bidder's support and by Lemma
  \ref{lem:another-bidder}, there are at least $2$ bidders such that
  $x$ is in their supports. 
	
	By the definition of derivative, for all
  $i\not\in S_x$, $g_{ij}(x)=0$. Similarly, we have $g_{ij}(x)>0$ and
  $G_{ij}(x)>0$ for all $i\in S_x$ by definition \ref{def:support}. 
	%By the definition of Nash equilibrium, if $g_{ij}(x)>0$, bidder $i$ must make her
  %equilibrium payoff when she bids $b_{ij}=x$, i.e. $x$ is good, if
  %$g_{ij}(x)>0$. 
	Moreover, for every $i \in S_x$, $x$ is good for bidder $i$ and item $j$, since $x$ is 
	in their support. 
	So, for any fixed $x\in (0, o_j-A_j],$ $\sum_{i\in[n]} F_{ij}(x)\cdot
  v_{ij}(x)\cdot g_{ij}(x)= \sum_{i\in S_x} F_{ij}(x)\cdot
  v_{ij}(x)\cdot g_{ij}(x),$ and according to Proposition
  \ref{prop:inequality},
  $$\sum_{i\in[n]} F_{ij}(x)\cdot v_{ij}(x)\cdot g_{ij}(x)\ge\sum_{i\in
    S_x}\frac{g_{ij}(x)}{\sum_{k\neq i, k\in S_x}\frac{g_{kj}}{G_{kj}}}\ge 
  \sqrt{\prod_{i\in S_x}G_{ij}(x)}\ge\sqrt{\prod_{i\in[n]}G_{ij}(x)} \enspace .$$ 
  Merging all these inequalities,
  $$SW(\B)\ge \sum_{j\in[m]}\int_0^{o_j-A_j}\sqrt{\prod_{i\in[n]}G_{ij}(x)}dx=\sum_{j\in[m]}\int_0^{o_j-A_j}\sqrt{F_j(x)}dx \enspace .$$
\end{proof}

\subsection{Proof of Inequality (\ref{eq:CDF-ineq})}
\label{sec:inequality-optimal-cdf}

In this section we prove the following technical lemma.
\begin{lemma}
  \label{lem:optimal-cdf} 
  For any CDF $F$ and any real $v>0$, $R(F,v)\ge \frac{3+4\lambda-\lambda^4}{6}v$.
\end{lemma}

In order to obtain a lower bound for $R(F,v)$ as stated in the lemma,
we show first that we can restrict attention to cumulative
distribution functions of a simple special form, since these
constitute worst cases for $R(F,v).$ % Observe
% that for functions $F$ with a given $A=\max_{x\geq 0}\{F(x)\cdot
% v-x\}$ it holds that $F(x)\leq \frac{x+A}{v}$ for any $x\geq 0.$
% In the next lemma, for an arbitrary CDF $F$ we will define a simple
% piecewise linear function $\hat F$ that has the same value of
% $\int_0^{v-A}(1-\hat F(x))dx=\int_0^{v-A}(1-F(x))dx$ as for $F,$ but (the same or) smaller
% value of $\int_0^{v-A}\sqrt{F(x)}dx$ than for $F.$ Once we establish
% this, it will be convenient to lower bound $R(\hat F,v)$ for the given
% type of piecewise linear functions $\hat F.$
%
In the next lemma, for an arbitrary CDF $F$ we will define a simple
piecewise linear function $\hat F$ that satisfies the following
two properties:
$$\int_0^{v-A}(1-\hat F(x))dx=\int_0^{v-A}(1-F(x))dx \; ; \;\; 
\int_0^{v-A}\sqrt{\hat F(x)}dx\leq \int_0^{v-A}\sqrt{F(x)}dx \enspace . $$ 

Once we establish this, it is convenient to lower bound $R(\hat
F,v)$ for the given type of piecewise linear functions $\hat F.$

\begin{lemma}
  \label{lem:restricted-form} 
  For any CDF $F$ and real $v>0$, 
  there always exists another CDF $\hat F$
  such that $R(F,v)\ge R(\hat F,v)$ that, for $A=\max_{x\geq 0}\{F(x)\cdot v - x\}$, is defined by
  \[\hat F(x)=\left\{
    \begin{array}{cl} 
      0  \enspace  & \mbox{, if } x \in [0, x_0] \\
      \frac{x+A}{v}  \enspace  & \mbox {, if } x \in (x_0, v-A] \enspace .
    \end{array}
  \right.\] 
\end{lemma}
\begin{proof}
  For any CDF $F$ and real $v>0$, 
  there always exists another CDF $\hat F$
  such that $R(F,v)\ge R(\hat F,v)$ that, for $A=\max_{x\geq 0}\{F(x)\cdot v - x\}$, is defined by
  \[\hat F(x)=\left\{
    \begin{array}{cl} 
      0 & \mbox{, if } x \in [0, x_0] \\
      \frac{x+A}{v} & \mbox { ,if } x \in (x_0, v-A] \enspace .
    \end{array}
  \right.\] 

\vspace{10pt}

First notice that $\max_{x\geq 0}\{\hat{F}(x)\cdot v - x\} = A$.
% so that the reader is not confused in the following, by wondering if $A$ is derived from  
% $F$ or $\hat{F}$.  
%  
By the definition of Riemann integration, we can represent the
integration as the limit of Riemann sums. For any positive integer
$l$, let $R_l$ be the Riemann sum if we partition the interval
$[0,v-A]$ into small intervals of size $(v-A)/l$. That is
  $$R_l(F,v)=A+\frac{v-A}{l}\cdot
  \left(\sum_{i=0}^{l-1}(1-F(x_i))+\lambda\cdot\sum_{i=0}^{l-1}\sqrt{F(x_i)}\right) \enspace ,$$
  where $x_i=\frac il \cdot (v-A)$. So we have
  $R(F,v)=\lim_{l\rightarrow \infty}R_l(F,v)$.
  
  For any given $l$, let $i^*$ be the index such that $\sum_{i>i^*}(x_i+A)/v <
  \sum_{i=0}^{l-1}F(x_i)$ and $\sum_{i>=i^*}(x_i+A)/v \ge
  \sum_{i=0}^{l-1}F(x_i)$. We define $\hat F_l$ as follows:
	
  \[\hat F_l(x)=\left\{
    \begin{array}{cl} 
      0 & \mbox{, if } x<x_{i^*}\\
      \sum_{i=0}^{l-1}F(x_i)-\sum_{i>i^*}(x_i+A)/v & \textrm {if } x\in[x_{i^*},x_{i^*+1})\\
      (x+A)/v & \mbox {, if } x\in[x_{i^*+1} ,v-A]  \enspace .
    \end{array}
  \right.\] 
	
	It is straight-forward to check that $\hat F(x) =\lim_{l\rightarrow \infty}\hat
  F_l(x)$, as described in the statement of the lemma. We will show that for
  any $l$, $R_l(F,v)\ge R_l(\hat F_l,v)$. Then the lemma follows by
  taking the limit, since $R_l(F,v)\rightarrow R(F,v),$ and $R_l(\hat
  F,v)\rightarrow R(\hat F,v).$ Figure \ref{fig:exampleLemma18}(a)
  illustrates $\hat F(x)$ (when we take the limit of $l$ to infinity).

  By the construction of $\hat F_l$, it is easy to check that
  $\sum_{i=0}^{l-1}F(x_i)=\sum_{i=0}^{l-1}\hat F_l(x_i)$ and
  $\max_x\{\hat F_l(x)\cdot v-x\}= A$. Then in order to prove $R_l(F,v)\ge
  R_l(\hat F_l,v)$, it is sufficient to prove that
  $\sum_{i=0}^{l-1}\sqrt{F(x_i)}\ge\sum_{i=0}^{l-1}\sqrt{\hat F_l(x_i)}$.
  Let $\mathcal{Q}$ be the set of CDF functions such that $\forall
  Q\in \mathcal{Q}$, $\sum_{i=0}^{l-1}Q(x_i)=\sum_{i=0}^{l-1}F(x_i)$
  and $A = \max_{x\geq 0} \{Q(x)\cdot v - x\}$, meaning further that 
	$Q(x)\leq (x+A)/v$, for all $x\geq 0$. We will show that $\hat F_l(x)$ has the minimum value for 
	the expression 
  $\sum_{i=0}^{l-1}\sqrt{\hat{F}_l(x_i)}$ within $\mathcal{Q}$. 
  \begin{figure}[h]
    \center
    \begin{tabular}{c c}
      \includegraphics[scale=0.5]{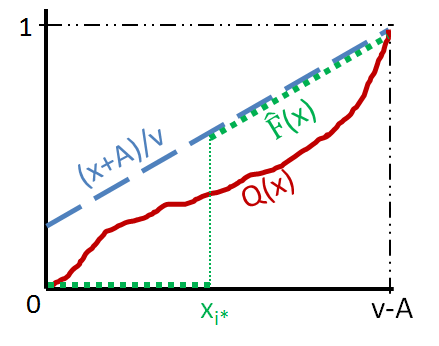} &
      \includegraphics[scale=0.5]{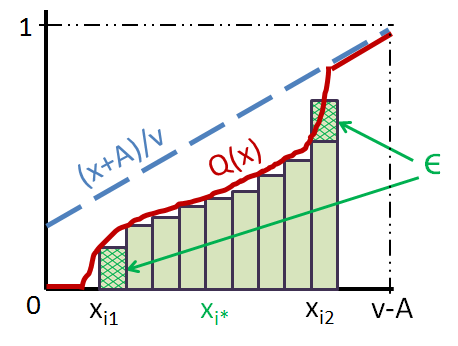}\\
      $(a)$&$(b)$
    \end{tabular}
    \caption{Figure $(a)$ illustrates $\hat F(x)=\lim_{l\rightarrow
        \infty}\hat F_l(x)$ and figure $(b)$ shows how $Q'$ is derived
      from $Q$.}
    \label{fig:exampleLemma18}
		\vspace{-10pt}
\end{figure}

Assume on the contrary that some other function $Q\in \mathcal{Q}$ has the minimum value for 
$\sum_{i=0}^{l-1}\sqrt{Q(x_i)}$ within $\mathcal{Q}$ and $Q(x_{j})\neq
\hat F_l(x_j)$ for some $x_j$. Let $i_1$ be the smallest index such
that $Q(x_{i_1})>0$ and $i_2$ be the largest index such that
$Q(x_{i_2})<(x_{i_2}+A)/v$. By the monotonicity of $Q$, we have $i_1\le
i_2$. Due to the assumption that $Q(x_j)\neq \hat F_l(x_j)$ for some
$x_j$ and $\sum_{i=0}^{l-1}\sqrt{Q(x_i)}\leq\sum_{i=0}^{l-1}\sqrt{\hat
  F_l(x_i)}$, we get $i_1\neq i_2$. So $i_1<i_2$ and
$Q(x_{i_1})<Q(x_{i_2})$ by the monotonicity of CDF functions. Now
consider another CDF function $Q'$ such that $Q'(x_i)=Q(x_i)$ for all
$i\neq i_1 \wedge i\neq i_2$, $Q'(x_{i_1})=Q(x_{i_1})-\epsilon$ and
$Q'(x_{i_2})=Q(x_{i_2})+\epsilon$ where $\epsilon = \min\{Q(x_{i_1}),
(x_{i_2}+A)/v-Q(x_{i_2})\}$. Figure \ref{fig:exampleLemma18}(b) shows
how we modify $Q$ to $Q'$. It is easy to check $Q'\in\mathcal{Q}$ and
$\sum_{i=0}^{l-1}\sqrt{Q(x_i)}>\sum_{i=0}^{l-1}\sqrt{Q'(x_i)}$ which
contradicts the optimality of $Q$. The inequality holds because of 
$\sqrt{a}+\sqrt{b}>\sqrt{a-c}+\sqrt{b+c}$ for all $0<c<a<b$, which can
be proved by simple calculations.
\end{proof}

Now we are ready to proceed with the proof of Lemma~\ref{lem:optimal-cdf}.
\begin{proof}[of Lemma \ref{lem:optimal-cdf}] By Lemma
  \ref{lem:restricted-form}, for any fixed $v>0$, we only need to
  consider the CDF's in the following form: for any positive
  $A$ and $x_0$ such that $x_0+A\le v$,
  \[F(x)=\left\{\begin{array}{cl} 
      0 & \mbox{, if } x \in [0, x_0] \\
      \frac{x+A}{v} & \mbox {, if } x \in (x_0, v-A] \enspace .
  \end{array}\right.\] 
Clearly, $\max_{x\geq 0}\{F(x)\cdot v-x\}=A$. Let $t=\frac{A+x_0}{v}$.  Then

\begin{eqnarray*}
  R(F,v)&=&A+\int_0^{v-A}(1-F(x))dx+\lambda\cdot\int_0^{v-A}\sqrt{F(x)}dx\\
  &=&A+v-A-\frac v2\cdot\left(\frac{x+A}{v}\right)^2\bigg\vert^{v-A}_{x_0}
  +\lambda\cdot\frac{2v}{3}\cdot \left(\frac{x+A}{v}\right)^{\frac 32}\bigg|^{v-A}_{x_0}\\
  &=&v-\frac v2\cdot(1-t^2)+\lambda\cdot\frac{2v}{3}\cdot(1-t^{\frac 32})
  = v\cdot\left(\frac 12(1+t^2)+\frac{2\lambda}{3}(1-t^{\frac 32})\right) \enspace .
\end{eqnarray*} 
By optimizing over $t$, the above formula is minimized when
$t=\lambda^2\le 1$. That is,
\[\quad\qquad\qquad R(F,v)\ge v\cdot\left(\frac
  12(1+\lambda^4)+\frac{2\lambda}{3}(1-\lambda^3)\right)=
	\frac{3+4\lambda-\lambda^4}{6}\cdot v \enspace . \qquad\quad\;\; \]
\end{proof}

% The bound 1.82 can be derived by combining these two lower bounds for
% $SW(\B)$ optimally as shown in Appendix \ref{sec:allpay-xos}. Finally,
% we show how to generalize the single-item proof to multi-item case. In
% order to do this, we introduce a new notion called expected
% marginal valuation denoted by $v_{ij}(x)$ which allows us to show that
% $F_{ij}(x)\cdot v_{ij}(x)-x \ge F_{ij}(y)\cdot v_{ij}(x)-y$. So we can
% apply above proof for each item separately and get the improved upper
% bound for simultaneous APAs.

%%% Local Variables:
%%% mode: latex
%%% TeX-master: "allpay"
%%% End:

%% file: multi-unit.tex
\section{Multi-unit Auctions}
\label{sec:multiUnit}
In this section, we propose a randomized all-pay mechanism for the
multi-unit setting, where $m$ identical items are to be allocated to
$n$ bidders. Markakis and Telelis~\cite{MT12} and de Keijzer
et~al.~\cite{KMST13} have studied the price of anarchy for
several multi-unit auction formats. The current best upper bound
obtained was 1.58 for both pure and mixed Nash equilibria. 

% The randomized PSAM is a randomized auction described as
% follows. Each bidder submits a single bid $b_i$ to the
% auctioneer.  After soliciting all the bids from the bidders, the
% auctioneer decides a (possibly rational) allocation $x_i$ for bidder $i$ equal to 
% $x_i=\frac{m\cdot b_i}{\sum_{i\in[n]}b_i}$ and payment $p_i=b_i$ if
% $\sum_ib_i>0$, otherwise $x_i=0$ and $p_i=0$ for all $i$.

We propose a {\em randomized} all-pay mechanism that induces a {\em
  unique pure} Nash equilibrium, with an improved price of anarchy bound of $4/3$.
We call the mechanism Random proportional-share allocation mechanism
(PSAM), as it is a randomized version of Kelly's celebrated
proportional-share allocation mechanism for divisible
resources~\cite{Kel97}. The mechanism works as follows (illustrated as 
Mechanism~\ref{alg:PSAM}).
%\begin{enumerate}
%\item Each bidder $i$ submits a single bid $b_i$ to the auctioneer.
%\item The auctioneer associates a real number $x_i=\frac{m\cdot
  %b_i}{\sum_{i\in[n]}b_i}$ with every bidder $i$. 
%\item Bidder $i$ is assigned $\down{x_i}$ items with certainty and one more item with probability $x_i-\down{x_i}$.
%\item Each player pays their bid, $p_i=b_i$\footnote{ In the degenerate case, where $\sum_ib_i=0$, then
  %$x_i=0$ and $p_i=0$ for all $i$.} for all $i$.
%\end{enumerate}

Each bidder submits a non-negative real $b_i$ to the auctioneer. After soliciting all the
bids from the bidders, the auctioneer associates a real number $x_i$
with bidder $i$ that is equal to $x_i=\frac{m\cdot
  b_i}{\sum_{i\in[n]}b_i}$. Each player pays their bid,
$p_i=b_i$. In the degenerate case, where $\sum_ib_i=0$, then
  $x_i=0$ and $p_i=0$ for all $i$.

We turn the $x_i$'s to a random allocation as follows. Each bidder $i$
secures $\down{x_i}$ items and gets one more item with probability
$x_i-\down{x_i}$. 
An application of the Birkhoff-von Neumann
decomposition theorem guarantees that given an allocation vector
$(x_1,x_2,\dots,x_n)$ with $\sum_ix_i=m$, one can always find a
randomized allocation\footnote{As an example, assume $x_1=2.5, x_2=1.6,
x_3=1.9$. One can define a random allocation such that assignments $(3,2,1)$,
$(3,1,2)$ and $(2,2,2)$  occur with probabilities $0.1$, $0.4$, and
$0.5$ respectively.} with random variables $X_1,X_2,\dots,X_n$ such
that $\E[X_i]=x_i$ and $\pr[\down{x_i} \leq X_i \leq \up{x_i}]=1$ (see for example
\cite{DFK11,CDW12}). 

We next show that the game induced by the Random PSAM when the
bidders have submodular valuations is {\em isomorphic} to the game induced
by Kelly's mechanism for a single divisible resource when bidders have piece-wise linear concave
valuations. For convenience, we review the definition of isomorphism between games
as appears in Monderer and Shapley~\cite{MS96}.

\begin{definition}
  \label{def:iso}
  {\em (Isomorphism \cite{MS96}).} Let $\Gamma_1$ and $\Gamma_2$ be games in strategic form with the
  same set of players $[n]$. For $k=1,2$, let $(A^i_k)_{i\in[n]}$ be
  the strategy sets in $\Gamma_k$, and let $(u^i_k)_{i\in[n]}$ be the
  utility functions in $\Gamma_k$. We say that $\Gamma_1$ and
  $\Gamma_2$ are isomorphic if there exists bijections
  $\phi^i:a^i_1\rightarrow a^i_2$, $i\in [n]$ such that for every $i\in
  [n]$ and every $(a^1,a^2,\dots, a^n)\in \times_{i\in[n]}A^i_1$,
  \[u^i_1(a^1,a^2,\dots,a^n)=u^i_2(\phi^1(a^1),\phi^2(a^2),\dots,\phi^n(a^n)) \enspace .\]
\end{definition}

\begin{algorithm}[tp]
  \SetAlgorithmName{Mechanism}{mechanism}{List of Mechanisms}
  \KwIn{Total number of items $m$ and all bidders' bid $b_1,b_2,\dots,b_n$}
  \KwOut{Ex-post allocations $X_1,X_2,\dots, X_n$ and payments
    $p_1,p_2,\dots, p_n$}
  \If{$\sum_{i\in[n]}b_i>0$}{
    \ForEach{bidder $i=1,2,\dots,n$}{
      $x_i\leftarrow \frac{m\cdot b_i}{\sum_{i\in[n]}b_i}$\;
      $p_i\leftarrow b_i$\;
    }
    Sample $\{X_i\}_{i\in[n]}$ from $\{x_i\}_{i\in[n]}$ by using
    Birkhoff-von Neumann decomposition theorem such that
    $\down{x_i}\le X\le\up{x_i}$ and the expectation of sampling
    $X_i$ is $x_i$\;
  }\lElse{
    Set $\vec{X}=\vec{0}$ and $\vec{p}=\vec{0}$\;
  }
    Return $X_i$ and $p_i$ for all $i\in [n]$\;
  \caption{Random PSAM}
  \label{alg:PSAM}
\end{algorithm}

\begin{figure}[h]
  \label{fig:concave}
  \begin{center}
    \includegraphics{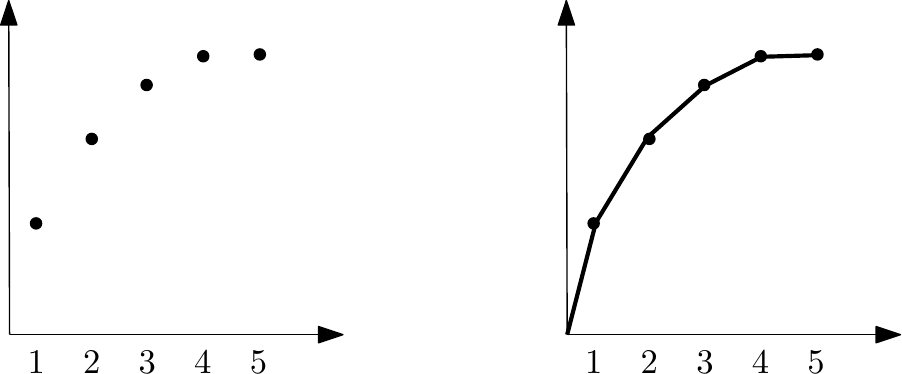}
  \end{center}
  \caption{The left part of the figure depicts some submodular
    function $f$, while the right part depicts the modified concave
    function $g$. One can verify that $g$ is concave if $f$ is submodular.}
\end{figure}

\begin{theorem}
  \label{thm:multi-unit}
  Any game induced by the Random PSAM applied to the multi-unit setting with
  submodular bidders is {\em isomorphic} to a game induced from
  Kelly's mechanism applied to a single divisible resource with piece-wise
  linear concave functions.
\end{theorem}
\begin{proof}
  For each bidder $i$'s submodular valuation function $f_i:\{0,1,\ldots,m\} \rightarrow R^+$, we associate
  a concave function $g_i:[0,1] \rightarrow R^+$ such that, 
	\begin{eqnarray}
	\forall x \in [0,m], \;\; g_i(x/m)=f_i(\down {x})+(x-\down
  {x})\cdot(f_i(\down {x}+1)-f_i(\down {x})) \enspace . \label{valuesEquiv}
	\end{eqnarray}
	Essentially, $g_i$ is the
  piecewise linear function that comprises the line segments that
  connect $f_i(k)$ with $f_i(k+1)$, for all nonnegative integers
  $k$. It is easy to see that $g_i$ is concave if $f_i$ is submodular
  (see also Fig. \ref{fig:concave} for an illustration).

  We use identity functions as the bijections $\phi^i$ of Definition
  \ref{def:iso}. Therefore, it suffices to show that, for any pure
  strategy profile $\b$, $u_i(\b)=u'_i(\b)$, where $u_i$ and $u'_i$
  are the bidder $i$'s utility functions in the first and second game,
  respectively. Let $x_i=\frac{m\cdot b_i}{\sum_ib_i}$, then
  \begin{eqnarray*}
	u_i(\b)&=&(x_i-\down {x_i})f_i(\down
  {x_i}+1) + (1-x_i+\down {x_i})f_i(\down {x_i})-b_i\\
	&=& f_i(\down {x_i}) + (x_i-\down {x_i})(f_i(\down {x_i}+1)-f_i(\down {x_i}))-b_i\\
	&=& g_i\left(\frac{x_i}{m}\right) - b_i = g_i\left(\frac{b_i}{\sum_ib_i}\right) - b_i = u'_i(\b) \enspace .
	\end{eqnarray*}
				Note that $g_i\left(\frac{b_i}{\sum_ib_i}\right) - b_i$ is player $i$'s utility, 
				under $\b$, in Kelly's mechanism. 
\end{proof}

We next show an equivalence between the optimal welfares. 
\begin{lemma}
\label{thm:opt-equal}
The optimum social welfare in the multi-unit setting, with submodular
valuations $\f = (f_1,\ldots, f_n)$, is equal to the optimal social
welfare in the divisible resource allocation setting with concave valuations
$\g = (g_1,\ldots g_n)$, where $g_i(x/m)=f_i(\down {x})+(x-\down
  {x})\cdot(f_i(\down {x}+1)-f_i(\down {x}))$.
\end{lemma}
\begin{proof}
    For any valuation profile $\v$ and (randomized) allocation 
	$\mathcal{A}$, we denote by $SW_{\v}(\mathcal{A})$ 
  the social welfare of allocation $\mathcal{A}$ under the
  valuations $\v$.
	For any fractional allocation $\x=(x_1,\ldots , x_n)$, such that 
	$\sum_ix_i = m$, let $\X(\x) = (X_1(\x),\ldots, X_n(\x))$ be the
  random allocation as computed by the Random PSAM given the
  fractional allocation $\x$.  Also let $\o=(o_1, \ldots , o_n)$ and
  $\O = (O_1, \ldots , O_n)$ be the optimal allocations in the
  divisible resource allocation problem and in the multi-unit auction,
  respectively. 
 
First we show that $SW_{\g}(\o) \geq SW_{\f}(\O)$. Consider the 
fractional allocation $\o' = (o'_1, \ldots , o'_n)$, where $o'_i = O_i/m$, for 
every $i$. Then it is easy to see that for every $i$, 
$g_i(o'_i) = f_i(\down {O_i})+(O_i-\down {O_i})\cdot(f_i(\down {O_i}+1)-f_i(\down {O_i})) = f_i({O_i})$, 
since $O_i$ is an integer. Therefore, $SW_{\g}(\o) \geq SW_{\g}(\o') = SW_{\f}(\O)$, 
by the optimality of $\o$. 

Now we show $SW_{\f}(\O) \geq SW_{\g}(\o)$. 
Note that for any fractional allocation $\x$, such that $\sum_jx_j = m$,  
$\E_{\X(\x)} [f_i({X_i(\x)})] = f_i(\down {x_i})+(x_i-\down {x_i})\cdot(f_i(\down {x_i}+1)-f_i(\down {x_i})) = g_i(x_i/m)$, for every $i$. 
By the optimality of $\O$, $SW_{\f}(\O) \geq \E_{\X(m\cdot\o)}[SW_{\f}(\X(m\cdot \o))] = SW_{\g}(\o)$.
\end{proof}

Theorem~\ref{thm:multi-unit} and Lemma~\ref{thm:opt-equal}, allow us
to obtain the existence and uniqueness of the pure Nash equilibrium,
as well as the price of anarchy bounds of Random PSAM by the
corresponing results on Kelly's mechanism for a single divisible
resource \cite{johari_efficiency_2004}. Moreover, it can be shown that
there are no other mixed equilibria by adopting the arguments of
\cite{CV14} for Kelly's mechanism. The main conclusion of this section
is summarized in the following Corollary.

\begin{corollary}
  % The social welfare of any mixed equilibrium in the mechanism using
  % above allocation and payment rules is at least $3/4$ of the optimal
  % social welfare.
Random PSAM induces a {\em unique pure} Nash equilibrium when applied to the
multi-unit setting with submodular bidders. Moreover, the price of
anarchy of the mechanism is {\em exactly} $4/3$.
\end{corollary}
% \begin{proof}
%   Theorem~\ref{thm:multi-unit} and \ref{thm:opt-equal}, the
%   existence and uniqueness of the pure Nash equilibrium and bounds for
%   PoA follows from the same results on Kelly's mechanism for single
%   indivisible item in \cite{johari_efficiency_2004}. Moreover, it can
%   be shown that there is no other mixed equilibria in this game by
%   adopting the same argument for Kelly's mechanism in \cite{CV14}.
% \end{proof}

%%% Local Variables:
%%% mode: latex
%%% TeX-master: "allpay"
%%% End:

%% file: singleItemPoA.tex
%\section{All-pay Contest}
\section{Single item auctions}
\label{sec:single}

In this section, we study mixed Nash equilibria in the single item
all-pay auction. 
First, we measure the inefficiency
of mixed Nash equilibria, showing tight results for the price of anarchy. 
En route, we also show that the price of anarchy is
$8/7$ for two players. Then we analyze the quality of two other
important criteria, the {\em expected revenue (the sum of bids)} and
the quality of the expected {\em highest submission (the maximum
  bid)}, which is a standard objective in crowdsourcing
contests~\cite{CHS12}.  For these objectives, we show a tight lower bound of
$v_2/2$, where $v_2$ is the second highest value among all bidders'
valuations. In the following, we drop the word expected while
referring to the revenue or to the maximum bid.

We quantify the loss of revenue and the highest submission in the
worst-case equilibria. We show that the all-pay auction achieves a
$2$-approximation comparing to the conventional procurement (modeled
as the first price auction), when considering worst-case mixed Nash
equilibria; we show in Appendix~\ref{firstPriceRevenue} that the
revenue and the maximum bid of the conventional procurement equals
$v_2$ in the worst case. We also consider other structures of rewards
allocation and conclude that allocating the entire reward to the
highest bidder is the only way to guarantee the approximation factor
of $2$. Roughly speaking, allocating all the reward to the top prize
is the optimal way to maximize the maximum bid and revenue among all
the prior-free all-pay mechanisms where the designer has no prior
information about the participants. Throughout this section we assume
that the players are ordered based on 
decreasing order of their valuations, i.e.  
$v_1 \geq v_2 \geq \ldots \geq v_n$. We also drop the word expected when
referring to the revenue or to the maximum bid.

\begin{theorem}
  \label{thm:PoA1.185} The mixed price of anarchy of the single item all-pay auction is
  $1.185$.
\end{theorem}
\begin{proof}
  {\bf Upper bound:}
  Based on the results of \cite{BKV96}, inefficient Nash equilibria
  only exist when players' valuations are in the form
  $v_1>v_2=...=v_k>v_{k+1}\geq ... \geq v_n$ (with $v_2>0$), where players $k+1$
  through $n$ bid zero with probability $1$. W.l.o.g., we assume that
  $v_1=1$ and $v_i=v>0$, for $2\leq i \leq k$. Let $P_1$ be the
  probability that bidder $1$ gets the item in any such mixed Nash
  equilibrium denoted by $\B$. Then the expected utility of bidder $1$
  in $\b\sim \B$ can be expressed by $\E[u_1(\b)]=P_1\cdot 1
  -\E[b_1]$. Based on the characterization in \cite{BKV96}, no player 
	would bid above $v$ in any Nash equilibrium and nobody bids exactly 
	$v$ with positive probability. Therefore, if player $1$ deviates to $v$, 
	she will gets the item with probability $1$. By the definition 
	of Nash equilibrium, we have $\E[u_1(\b)] \geq \E[u_1(v,\b_{-i})]=1-v$, 
	resulting in $P_1 \geq 1-v + \E[b_1]$.

  It has been shown in the proof of Theorem $2C$ in \cite{BKV96}, that
  $\E[b_1]$ is minimized when players $2$ through $k$ play symmetric
  strategies. Following their results, we can extract the following
  equations (for a specific player $i$):
  \[ G_1(x)=\frac{x}{v \prod_{i'\neq 1,i}{G_{i'}(x)}},
    \quad \forall x \in (0,v] \enspace ; \quad \;\;
     \prod_{i'\neq 1}{G_{i'}(x)} = 1-v+x, \quad \forall x \in
    (0,v] \enspace . \]
  Recall that $G_{i'}(x)$ is the CDF according to which player $i'$ bids in
  $\B$. Since players $2$ through $k$ play symmetric
  strategies, $G_{i'}(x)$ should be identical for $i'\neq 1$. Then, for some $i' \neq 1$, 
  $$ G_1(x)=\frac{x}{v\cdot G_{i'}^{k-2}(x)}=\frac{x}{v\cdot
    \left(1-v+x\right)^{\frac{k-2}{k-1}}} \enspace , \qquad \forall x \in
  (0,v] \enspace .\quad\qquad\qquad\qquad\qquad\quad$$
  Note that $1-v+x \leq 1$, and so we get $G_1(x) \leq \frac{x}{v \left(1-v+x\right)}$
  (for two players, $G_1(x) = \frac{x}{v }$) and 
  $$\E[b_1] \geq \int_0^v \left(1-\frac{x}{v
        \left(1-v+x\right)}\right)dx = v-1-\frac{(1-v)\ln
      (1-v)}{v} \enspace .\qquad\qquad\qquad\quad\;$$
  Now we can derive that 
	$P_1 \geq \frac{1-v}{v}\ln \frac{1}{1-v}.$\\ %\qquad\qquad\qquad\qquad\qquad\qquad\qquad\qquad\qquad\qquad\qquad$$ 
	For two players, $\E[b_1]=\int_0^v \left(1-x/v \right)dx=v/2$ and so
  $P_1 = 1-v/2$. 
	
	The expected social welfare in $\B$ is
  $\E[SW(b)] \geq P_1+(1-P_1)v \geq \frac{(1-v)^2}{v}\ln \frac{1}{1-v} + v$.
  The expression, $T(v) = \frac{(1-v)^2}{v}\ln \frac{1}{1-v} + v$, is
  minimized for $v\approx 0.5694$ and therefore, the price of anarchy is at most $T(0.5694) \approx
  1.185$. Particularly, for two players, $\E[SW(b)] \geq 1-v/2+v^2/2$,
  which is minimized for $v=1/2$ and therefore the price of anarchy for two players
  is at most $8/7$.

\vspace{5pt}

{\bf Lower bound:}
  Consider $n$ players, with valuations $v_1=1$ and $v_i=v>0$, for
  $2\leq i\leq n$. Let $\B$ be the Nash equilibrium, where bidders bid
  according to the following CDFs, 
  $$G_1(x)=\frac{x}{v \left(1-v+x\right)^\frac{n-2}{n-1}} \; x\in [0,v] \enspace ; \;\;  
	G_i(x)=\left(1-v+x\right)^\frac{1}{n-1} \; x\in [0,v],\; i\neq 1 \enspace . $$
  Note that $F_i(x) = \prod_{i'\neq i} G_{i'}(x)$ is the probability of
  bidder $i$ getting the item when she bids $x$, for every bidder $i$.
  $$F_1(x) = (1-v+x)\qquad x\in [0,v] \enspace ; \qquad \qquad F_i(x)=\frac{x}{v} \qquad x\in [0,v], \;\;\;i \neq 1 \enspace .\quad$$
  If player $1$ bids any value $x \in [0,v]$, her utility is $u_1 =
  F_1(x) \cdot 1 -x = 1-v$. Bidding greater than $v$ is dominated by
  bidding $v$. If any player $i\neq 1$ bids any value $x\in [0,v]$,
  her utility is $u_i = F_i(x) \cdot v -x = 0$. Bidding greater than
  $v$ results in negative utility. Hence, $\B$ is a Nash equilibrium.
  Let $P_1$ be the probability that bidder $1$ gets the item in $\B$, then
  $$ \E[SW(b)] = 1\cdot P_1+(1-P_1)v = v+(1-v)P_1
    = v+(1-v)\int_0^v G_i^{n-1}(x) dG_1(x) \enspace .$$
  When $n$ goes to infinity, $\E[SW(b)]$ converges to $v+(1-v)\int_0^v
  \frac{1-v}{v(1-v+x)}dx =
  v+(1-v)\frac{1-v}{v}\ln{\frac{1}{1-v}}=\frac{(1-v)^2}{v}\ln
  \frac{1}{1-v} + v = T(v)$. If we set $v=0.5694$, the price of anarchy is at least $T(v) \approx 1.185$. \\
  For $n=2$, $\E[SW(b)] = v+(1-v) \int_0^v\frac{1-v+x}{v} =
  v+(1-v)(1-v/2) = 1-v/2+v^2/2$, which for $v=1/2$ results in price of anarchy at
  least $8/7$.
\end{proof}

%\subsection{Revenue and Maximum Bid}
%\label{sec:single-max-bid}
%In this section we bound the revenue and the maximum bid of the
%single-item all-pay auction, for the case of mixed Nash
%equilibria. Specifically, for $v_1$ and $v_2$ being the highest and the second highest 
%valuations respectively, the revenue and the maximum bid have value 
%of at least $v_2/2$ and this value goes to
%$v_2/2$ when the number of bidders goes to infinity and $v_2/v_1$
%approaches $0$.

% We lower bound both values by one half of the
% second highest valuation, meaning that the expected revenue and
% maximum bid of all-pay auction approximate by $1/2$ the worst-case
% corresponding values of the first-price auction (or under the
% no-overbidding assumption).

\begin{theorem}
\label{maxBidAllPay}
In any mixed Nash equilibrium of the single-item all-pay auction, the
revenue and the maximum bid are at least half of the second highest
valuation.
\end{theorem}
\begin{proof}
    Let $k$ be any integer greater or equal to $2$, such that $v_1 \geq
  v_2 = \ldots = v_k \geq v_{k+1} \geq \ldots \geq v_n$. Let $F(x) =
  \prod_i G_i(x)$ be the CDF of the maximum bid $h$. By the
  characterization of \cite{BKV96}, in any mixed Nash equilibrium,
  players with valuation less than $v_2$ do not participate (always
  bid zero) and there exist two players $1, i$ bidding continuously in
  the interval $[0,v_2]$.  Then, by \cite{BKV96}, $F_1 =
  (v_1-v_2+x)/{v_1}$ and $F_{i}(x)=x/ v_2$, for any $x \in
  (0,v_2]$. Therefore, we get
$$ F(x) = F_{i}(x) G_{i}(x)= \frac{x}{v_2} G_{i}(x) \enspace .$$

In the proof of Theorem 2C in \cite{BKV96}, it is argued that
$G_{i_1}(x)$ is maximized (and therefore the maximum bid is
minimized) when all the $k$ players play symmetrically (except for
the first player, in the case that $v_1 > v_2$). So, $F(x)$ is
maximized for $G_{i}= \left(\prod_{i'\neq 1}
  G_{i'}\right)^{\frac{1}{k-1}} = F_1^{\frac{1}{k-1}} =
\left(\frac{v_1-v_2+x}{v_1} \right)^{\frac{1}{k-1}}$.
 Finally we get
\begin{eqnarray*} 
E[h] &=& \int_0^\infty (1-F(x))dx \geq \int_0^{v_2}\left(1-\frac{x}{v_2}\left(\frac{v_1-v_2+x}{v_1} \right)^{\frac{1}{k-1}}\right) dx \\
&\geq & v_2 - \int_0^{v_2}\frac{x}{v_2}dx = \frac 12 v_2 \enspace . 
\end{eqnarray*} 

The same lower bound also holds for the revenue, which is at
least as high as the maximum bid. This lower bound is tight for the maximum bid, as
indicated by our analysis, when $k$ goes to infinity and for the symmetric mixed 
Nash equilibrium. In the next lemma, we show that this lower bound is also tight 
for the revenue. 
\end{proof}

\begin{lemma}
\label{revAllPay}
For any $\epsilon>0$, there exists a valuation vector
$\v=(v_1,\dots,v_n)$, such that in a mixed Nash equilibrium of the
induced single-item all-pay
auction, the revenue and the maximum bid is at most $v_2/2+\epsilon$.
\end{lemma}
\begin{proof}
    In \cite{BKV96}, the authors provide results for the revenue in all
  possible equilibria. For the case that $v_1 = v_2$, the 
  revenue is always equal to $v_2$. To show a tight lower bound, we
  consider the case where $v_1 > v_2$ and there exist $k$ players with
  valuation $v_2$ playing symmetrically in the equilibrium, by letting $k$
  go to infinity. For this case, based on \cite{BKV96}, the revenue
  is equal to\footnote{For simplicity we assume $v_1=1$ and $v_2=v$.}

$$ \sum_i\E[b_i] = v^2+\left(1-{v}\right)\E[b_1] \enspace ,$$
where $\E[b_1] = \int_0^{v} \left(1-G_1(x)\right)dx$. From the proof of Theorem 
\ref{maxBidAllPay} we can derive that $G_1(x) = F(x)/F_1(x) = 
\frac{x}{v}\left({1-v+x} \right)^{\frac{1}{k-1}-1} = 
\frac{x}{v}\left({1-v+x} \right)^{-1}$, when $k$ goes to infinity. By
substituting we get,
\begin{eqnarray*}
\sum_i\E[b_i] &=& v^2+\left(1-{v}\right)
\int_0^{v}\left(1-\frac{x}{v}\left({1-v+x} \right)^{-1}\right)dx\\
&=& v^2+\left(1-{v}\right)\left(v-\frac{1}{v}
\left(v + (1-v)\ln(1-v)\right)\right)\\
&=& 2v - 1 - \frac{(1-v)^2}{v}\ln(1-v)\\
&=& v - (1-v)\left(1 + \frac{1-v}{v}\ln(1-v)\right) \enspace .
\end{eqnarray*}  
By taking limits, we finally derive that 
$\lim_{v \rightarrow 0} \left(\frac{\sum_i\E[b_i]}{v}\right) = 1/2$. 
The same tightness result also holds for the maximum bid, which is at
most the same as the revenue.
\end{proof}

Finally, the next theorem indicates that allocating the entire reward
to the highest bidder is the best choice. In particular a prior-free
all-pay mechanism is presented by a probability vector $\q= (q_i)_{i\in[n]}$,
with $\sum_{i\in[n]} q_i = 1$, where $q_i$ is the probability that the
$i^{th}$ highest bidder is allocated the item, for every $i \leq n$.
% any prior-free mechanisms that does not
% allocate the entire reward to the highest bidder, can
% achieve a revenue and quality of the best bid that is strictly worse
% than $v_2/2$ in all mixed Nash equilibria.

% Finally, we show that not allocating the entire reward to
% the bidder with the highest bid cannot guarantee that the revenue and the quality of the
% best bid is at least $v_2/2$ in all mixed Nash equilibria. 

%More specifically, we consider any probability vector $\q = (q_i)_i$, where 
%$\sum_i q_i = 1$ and $q_i$ is the probability that the $i^{th}$ 
%highest bidder is allocated the reward, for every $i \leq n$. 
%Similar to \cite{CHS12}, we make the assumption that $\q$ consists 
%of decreasing values. 

\begin{theorem}
\label{thm:allToHighest}
  For any prior-free all-pay mechanism that assigns the item to the
  highest bidder with probability strictly less than $1$,
  i.e. $q_1<1$, there exists a valuation profile and mixed Nash
  equilibrium such that the revenue and the maximum bid are
  strictly less than $v_2/2$.
\end{theorem}
\begin{proof}
     We will assert the statement of the theorem for the valuation profile $(1, v,
  0,0,\dots,0)$, where $v\in (0,1)$ is the second highest value. It is
  safe to assume that $q_2\in [0,q_1)$ \footnote{Otherwise, consider
    the tie-breaking rule that allocates the item equiprobably. Then
    for $q_2 \geq q_1$, the strategy profile where all players bid
    zero is strictly dominant.}.  We show that the following bidding
  profile is a mixed Nash equilibrium.  The first two bidders bid on
  the interval $[0,v(q_1-q_2)]$ and the other bidders bid $0$.  The
  CDF of bidder $1$'s bid is $G_1(x)=\frac{x}{v(q_1-q_2)}$ and the CDF
  of bidder $2$'s bid is $G_2(x)=x/(q_1-q_2)+1-v$. It can be checked
  that this is a mixed Nash equilibrium by the following
  calculations. For every bid $x\in [0,v(q_1-q_2)]$,
  \[u_1(x)=G_2(x)\cdot q_1+(1-G_2(x))\cdot q_2-x= q_1-v(q_1-q_2) \enspace ,\]
  \[u_2(x)=G_1(x)\cdot q_1v+(1-G_1(x))\cdot q_2v-x= q_2v \enspace .\]
  %The maximum bid is 
  %\begin{align*}
  %\int_{x=0}^{v(q_1-q_2)}(1-G_1(x)G_2(x))dx&=\int_{x=0}^{v(q_1-q_2)}\left(1-\frac{x}{v(q_1-q_2)}\cdot
  %\left(\frac{x}{q_1-q_2}+1-v\right)\right)dx\\
  %&=\frac{v(q_1-q_2)}{2}+\frac{v^2(q_1-q_2)}{6}
  %\end{align*}
	The revenue is 
	\begin{align*}
  &\int_{0}^{v(q_1-q_2)}(1-G_1(x))dx + \int_{0}^{v(q_1-q_2)}(1-G_2(x))dx\\
	=&\int_{0}^{v(q_1-q_2)}\left(1-\frac{x}{v(q_1-q_2)}\right)dx 
	+ \int_{0}^{v(q_1-q_2)}\left(1-\left(\frac{x}{q_1-q_2}+1-v\right)\right)dx\\
  =&\frac{v(q_1-q_2)}{2}+\frac{v^2(q_1-q_2)}{2} \enspace .
  \end{align*}
	When $v$ goes to $0$, the revenue go to $v(q_1-q_2)/2<v/2$ since $q_1-q_2<1$. 
	Obviously, the same happens with the maximum bid, which is at most the same as 
	the revenue. 
\end{proof}

%%% Local Variables: 
%%% mode: latex
%%% TeX-master: "allpay"
%%% End: 

%% file: appendixXosInequality.tex
\section{Proof of Proposition \ref{prop:inequality}}
\label{sec:inequality}

\textsc{Proposition \ref{prop:inequality} (restated).} 
  For any integer $n\ge 2$, any positive real $G_i\le 1$ and positive
  real $g_i$ for $1\leq i\leq n,$
  $$\sum_{i=1}^n\frac{g_i}{\sum_{k\neq i}\frac{g_k}{G_k}}
  \ge\sqrt{\prod_{i=1}^nG_i}\enspace .$$
	
	\vspace{10pt}
	
In order to prove the proposition, we will minimize the left hand side of the inequality over all  $G_i$
and $g_i$,  such that
\begin{equation}
  \label{eq:distCnstr}
  0<G_i\le 1\quad\quad g_i>0 \quad (i\in [n]) \quad \mbox{ where} \quad \prod_{t=1}^n
  G_t \quad \mbox{is a constant}  \enspace .
\end{equation}
We introduce the following notation:
$$H = \sum_{i=1}^n \frac{g_i}{\sum_{t=1,t\neq i}^n\frac{g_t}{G_t}}
\qquad \mbox{and} \qquad \forall i , \qquad H_i =
\frac{g_i}{\sum_{t=1,t\neq i}^n\frac{g_t}{G_t}} \enspace .$$
Note that $H = \sum_{i=1}^n H_i.$ Our goal is to minimize  $H$ over all possible variables $G_i$ and $g_i$ under the constraints
\eqref{eq:distCnstr}, and eventually show $H\ge\sqrt{\prod_{i=1}^nG_i}$. We also use the notation  $\G = (G_i)_i,$  $\g = (g_i)_i$, $H =
H(\G,\g)$ and $H_i = H_i(\G,\g)$, $\forall i$.

\begin{lemma}
\label{formOfOPt}
For every $\G$ and $\g$ that minimize $H(\cdot,\cdot)$ under constraints \eqref{eq:distCnstr}:
\begin{enumerate}
\item If $G_i < 1$ and $G_j < 1$, then $H_i=H_j,$
\item If $G_i = G_j = 1$ then $g_i = g_j$.
\end{enumerate}
\end{lemma}

We prove Lemma \ref{formOfOPt}, by proving Lemmas \ref{equalTerms} and \ref{equal_g_forG=1}.

\begin{lemma}
\label{equalTerms}
Under constraints \eqref{eq:distCnstr}, if $\G$ and $\g$ minimize $H(\cdot,\cdot)$, then for every $G_i < 1$ and $G_j <1$, $H_i(\G,\g) = H_j(\G,\g)$. 
\end{lemma}

\begin{proof}
For the sake of contradiction, suppose that there exist $G_i < 1$ and $G_j <1$ such that (w.l.o.g.) $H_i(\G,\g) > H_j(\G,\g)$. 
%We will iteratively change $\G$ and $\g$, until we end up with $\G'$ and $\g'$ as in lemma. At each iteration either some $G_i<1$ is set to $1$ or %we strictly reduce the value of some $H_i=\max_s(H_s|G_s <1)$ (if there are more than one such $i$, we need that many iterations in order to strictly reduce the value of $\max_s(H_s|G_s <1)$). 
%$H_i(\G',\g') = H_j(\G',\g')$. Moreover, at each iteration $H(\G,\g)> H(\G',\g')$ which guarantees that the procedure will finish. In the following we demonstrate a single iteration.
%W.l.o.g. suppose that for $G_i < 1$ and $G_j <1$, $H_i(\G,\g) > H_j(\G,\g)$. %and choose $i$ such that $H_i(\G,\g)$ is the {\em maximum possible} (i.e. $H_i(\G,\g)$ = $\max_s(H_s(\G,\g)|G_s <1)$).
Let $$r=\min\left\{\left(\frac{H_i(\G,\g)}{H_j(\G,\g)}\right)^{\frac 12}, \frac{1}{G_j}\right\} \enspace .$$ 
Notice that $r > 1$.

{\em Claim}: We claim that $H(\G,\g) > H(\G',\g')$, where $\G' = (\frac{G_i}{r}, rG_j, \G_{-ij})$ and $\g'= (\frac{g_i}{r}, rg_j, \g_{-ij})$.\\ 
As usual $\G_{-ij}$ stands for $\G$ vector after eliminating $G_i$ and $G_j$ (accordingly for $\g_{-ij}$). Therefore $\G'$ and $\g'$ are the same as $\G$ and $\g$ by replacing $G_i$, $G_j$, $g_i$, $g_j$ by $\frac{G_i}{r}$, $rG_j$, $\frac{g_i}{r}$, $rg_j$, respectively.

{\em Proof of the claim}: Notice that 
$$\frac{g'_i}{G'_i}=\frac{g_i/r}{G_i/r}=\frac{g_i}{G_i}  \enspace , \;\; \frac{g'_j}{G'_j}=\frac{rg_j}{rG_j}=\frac{g_j}{G_j} \quad \mbox{and} \quad \forall s \neq i,j , \;\; G'_s = G_s \;\; \mbox{and} \;\; g'_s = g_s \enspace .$$
Therefore, $\forall s \neq i,j$, $H_s(\G,\g)=H_s(\G', \g')$. So, we only need to show that $H_i(\G,\g)+H_j(\G,\g) > H_i(\G',\g')+H_j(\G',\g')$.

\begin{eqnarray*}
&& H_i(\G',\g')+H_j(\G',\g')\\
&=&\frac{g'_i(x)}{\sum_{t=1,t\neq i}^n\frac{g'_t(x)}{G'_t(x)}} + \frac{g'_j(x)}{\sum_{t=1,t\neq j}^n\frac{g'_t(x)}{G'_t(x)}}\\
&=&\frac{g_i(x)/r}{\sum_{t=1,t\neq i}^n\frac{g_t(x)}{G_t(x)}} + \frac{rg_j(x)}{\sum_{t=1,t\neq j}^n\frac{g_t(x)}{G_t(x)}}\\
&=& \frac{H_i(\G,\g)}{r} + rH_j(\G,\g)\\
&=& \left(\frac{1}{r}-1\right)H_i(\G,\g) + (r-1)H_j(\G,\g) + H_i(\G,\g)+H_j(\G,\g)\\
&\leq& \left(\frac{1}{r}-1\right)r^2H_j(\G,\g) + (r-1)H_j(\G,\g) + H_i(\G,\g)+H_j(\G,\g)\\
&=& -\left(r-1\right)^2H_j(\G,\g) + H_i(\G,\g)+H_j(\G,\g)\\
&<& H_i(\G,\g)+H_j(\G,\g) \enspace .
\end{eqnarray*}

In the above inequalities we used that $r>1$ and $r^2\leq
\frac{H_i(\G,\g)}{H_j(\G,\g)}$. The claim contradicts the assumption
that $H(\G,\g)$ is the minimum, so the lemma holds. 
\end{proof}

\begin{lemma}
\label{equal_g_forG=1}
Under constraints \eqref{eq:distCnstr}, if $\G$ and $\g$ minimize $H(\cdot,\cdot)$, then for every $G_i = G_j =1$, $g_i=g_j.$
%?? changed from $g'_i=g'_j$. 
\end{lemma}

\begin{proof}
For the sake of contradiction, suppose that there exist $G_i = G_j =1$ such that $g_i \neq g_j$. 
We will prove that for $\g' = (\frac{g_i+g_j}{2},\frac{g_i+g_j}{2},g_{-ij})$ (i.e. for every $k \neq i , j$, $g'_k = g_k$, and $g'_i=g'_j= \frac{g_i+g_j}{2}$), $H(\G,\g)> H(\G,\g')$. 
%By iteratively modifying $\g$, the iterations will end at some $\g'$, such that the lemma holds. The fact that at each iteration $H$ strictly decreases guarantee the end of the procedure.

Notice that for every $k \neq i , j$, $H_k(\G,\g') = H_k(\G,\g)$, since $g_i+g_j=g'_i+g'_j$ and $G_i = G_j =1$. Hence it is sufficient to show that $H_i(\G,\g)+H_j(\G,\g) \geq H_i(\G,\g')+H_j(\G,\g')$. 
Let $A_{ij} = \sum_{t \neq j,t\neq i}\frac{g_t}{G_t}$.

\begin{eqnarray*}
&&H_i(\G,\g) + H_j(\G,\g) - H_i(\G,\g') - H_j(\G,\g')\\
&=& \frac{g_i}{g_j + A_{ij}} +\frac{g_j}{g_i + A_{ij}} - \frac{g_i}{\frac{g_i+g_j}{2} + A_{ij}} - \frac{g_j}{\frac{g_i+g_j}{2} + A_{ij}}\\
&=& \frac{g_i}{g_j + A_{ij}} +\frac{g_j}{g_i + A_{ij}} - \frac{2g_i+2g_j}{g_i+g_j + 2A_{ij}} \\
&=&g_i\frac{(g_i+A_{ij})((g_i+g_j+2A_{ij}) - 2(g_j+A_{ij}))}{(g_j + A_{ij})(g_i + A_{ij})(g_i+g_j + 2A_{ij})}\\
&+&g_j\frac{(g_j+A_{ij})((g_i+g_j+2A_{ij}) - 2(g_i+A_{ij}))}{(g_j + A_{ij})(g_i + A_{ij})(g_i+g_j + 2A_{ij})}\\
&=&\frac{g_i(g_i+A_{ij})(g_i-g_j)+g_j(g_j+A_{ij})(g_j-g_i)}{(g_j + A_{ij})(g_i + A_{ij})(g_i+g_j + 2A_{ij})}\\
&=&\frac{(g_i-g_j)(g^2_i - g_j^2 + A_{ij}(g_i-g_j))}{(g_j + A_{ij})(g_i + A_{ij})(g_i+g_j + 2A_{ij})}\\
&=&\frac{(g_i-g_j)^2(g_i + g_j + A_{ij})}{(g_j + A_{ij})(g_i + A_{ij})(g_i+g_j + 2A_{ij})} > 0 \enspace ,
\end{eqnarray*}
which contradicts the assumption that $\G$ and $\g$ minimize $H(\cdot,\cdot)$. 
\end{proof}

%?? I changed to >0

\begin{lemma}
\label{relation_gG}
If $H_i=H_j$, then:
\begin{enumerate}
\item $g_i=g_j \Leftrightarrow G_i=G_j$, 
\item $(g_i=rg_j>0 \mbox{ and } r\geq 1) \Rightarrow G_i \geq r^2G_j$.
\end{enumerate} 
\end{lemma}

\begin{proof}
Let $A_{ij} = \sum_{t \neq j,t\neq i}\frac{g_t}{G_t}$; then $H_i = \frac{g_i}{\frac{g_j}{G_j}+A_{ij}}$. By assumption:
\begin{eqnarray*}
\frac{g_i}{\frac{g_j}{G_j}+A_{ij}} &=& \frac{g_j}{\frac{g_i}{G_i}+A_{ij}}\\
\frac{g_i^2}{G_i} + g_iA_{ij} &=& \frac{g^2_j}{G_j} + g_jA_{ij}\\
(g_i- g_j)A_{ij} &=&  \frac{g^2_j}{G_j} - \frac{g_i^2}{G_i} \enspace .
\end{eqnarray*}
If $g_i = g_j$ then $\frac{1}{G_j} - \frac{1}{G_i}=0$, so $G_i=G_j$. \\
If $G_i = G_j$ then $(g_i-g_j)(g_i+g_j + A_{ij}G_i) = 0$ . Under constraints \eqref{eq:distCnstr}, 
$A_{ij}G_i > 0$ and $g_i,g_j> 0$, so $g_i-g_j=0$ which results in $g_i=g_j$. \\
If $g_i=rg_j$, with $r\geq 1$ then $(g_i- g_j)A_{ij} \geq 0$ and so $\frac{1}{G_j} - \frac{r^2}{G_i} \geq 0$, which implies $G_i \geq r^2G_j.$ 
\end{proof}

\begin{lemma}
\label{optimize_g} 
For $n,k$ integers, $n\geq 2$, $1\leq k \leq n$, $0< a\leq 1$ and $g>0$:
$$L=\frac{kg}{(k-1)\frac{g}{a}+n-k}+\frac{n-k}{k\frac{g}{a}+n-k-1}\geq a \enspace .$$
\end{lemma}

\begin{proof}
We distinguish between two cases, 1) $k > \frac{1}{1-\sqrt{a}}$ and 2) $k \leq \frac{1}{1-\sqrt{a}}$.

\noindent{\em Case 1} ($k > \frac{1}{1-\sqrt{a}}$): For $k=n$,  $L = \frac{k}{k-1}a \geq a$. 
We next show that $\frac{dL}{dg} \leq 0$, for $n\geq 2$, $1\leq k < n$, $0< a\leq 1$ and $g>0$.
\begin{eqnarray*}
\frac{dL}{dg} = \frac{(n-k)k}{\left(\frac{(k-1)g}{a}+n-k\right)^2}-\frac{(n-k)k}{\left(\frac{kg}{a}+n-k-1\right)^2 a} &\leq& 0\\
\left(\frac{(k-1)g}{a}+n-k\right)^2 - \left(\frac{kg}{a}+n-k-1\right)^2 a &\geq& 0\\
\left(\frac{(k-1)g}{a}+n-k  +\left( \frac{kg}{a}+n-k-1\right)a^\frac 12\right) \qquad\qquad\qquad\qquad\qquad\qquad &&\\ 
\cdot \left(\frac{(k-1)g}{a}+n-k - \left( \frac{kg}{a}+n-k-1\right)a^\frac 12\right) &\geq& 0\\
\left(\frac{(k-1)g}{a}+n-k - \left( \frac{kg}{a}+n-k-1\right)a^\frac 12\right) &\geq& 0\\
\left(\frac{g}{a}\left(k-1-ka^\frac 12\right)+(n-k)\left( 1-a^\frac 12\right) +a^\frac 12\right) &\geq& 0\\
k-1-ka^\frac 12 &\geq& 0  \enspace ,
\end{eqnarray*}
which is true by the case assumption. Therefore, $L$ is non-increasing and so it is minimized for $g=\infty$. Hence, $L \geq \frac{k}{k-1}a \geq a$.

\noindent{\em Case 2} ($k \leq \frac{1}{1-\sqrt{a}}$): $L$ is minimized ($dL/dg (g^*) = 0$) 
for $g^* = \frac{a\left( \sqrt{a}+(n-k)\left( 1-\sqrt{a} \right) \right)}{k\sqrt{a}-k+1}$, therefore:
$$L \geq \frac{k^2\left(1-\sqrt{a}\right)^2+k\left(a-n\left(1-\sqrt{a}\right)^2-1\right)+n)}{(n-1)} \enspace ,$$
which is minimizes for $k = \frac{n}{2}+\frac{\left(1+\sqrt{a}\right)}{2\left(1-\sqrt{a}\right)}$. 
However, for $n\geq 2$, $\frac{n}{2}+\frac{\left(1+\sqrt{a}\right)}{2\left(1-\sqrt{a}\right)} \geq \frac{1}{1-\sqrt{a}}$. 
Notice, though, that for $k \leq \frac{1}{1-\sqrt{a}}$, $L$ is decreasing, so it is minimized 
for $k = \frac{1}{1-\sqrt{a}}$. Therefore, $L \geq \sqrt{a} \geq a$. 
\end{proof}

\vspace{10pt}

\begin{proof}
(Proposition \ref{prop:inequality})\\
%If there exists $i$ such that $G_i = 0$ then Proposition \ref{prop:inequality} follows. 
%Therefore we can assume constraints \eqref{eq:distCnstr}.\\
 Let $\G$ and $\g$ minimize $H(\cdot,\cdot)$ and also let $S=\{i|G_i<1\}$ and $F=\prod_{t=1}^n G_t$. Moreover, given Lemma \ref{formOfOPt}, for $g_i=\hat{g}$ for every $i \notin S$ and $j=\arg \min_{i\in S}g_i$, $H(\G,\g)$ can be written as:
$$H(\G,\g) = |S|\frac{g_j}{\sum_{t \in S, t \neq j}\frac{g_t}{G_t}+(n-|S|)\hat{g}}+(n-|S|)\frac{\hat{g}}{\sum_{t \in S}\frac{g_t}{G_t}+(n-|S|-1)\hat{g}} \enspace .$$
Let $g_i = r_ig_j$, for every $i \in S$. Since $j=\arg \min_{i\in S}g_i$, then for every $i \in S$, $r_i \geq 1$. By using Lemma \ref{relation_gG}:

\begin{eqnarray*}
H(\G,\g) &=& \frac{|S|\cdot g_j}{\sum_{t \in S, t \neq j}\frac{r_tg_j}{G_t^\frac 12 G_t^\frac 12}+(n-|S|)\hat{g}}+\frac{(n-|S|)\cdot \hat{g}}{\sum_{t \in S}\frac{r_tg_j}{G_t^\frac 12 G_t^\frac 12}+(n-|S|-1)\hat{g}}\\
&\geq& \frac{|S|\cdot g_j}{\sum_{t \in S, t \neq j}\frac{r_tg_j}{(r_t^2G_j)^\frac 12 G_t^\frac 12}+(n-|S|)\hat{g}}+\frac{(n-|S|)\cdot \hat{g}}{\sum_{t \in S}\frac{r_tg_j}{(r_t^2G_j)^\frac 12 G_t^\frac 12}+(n-|S|-1)\hat{g}}\\
&\geq& \frac{|S|\cdot g_j}{\sum_{t \in S, t \neq j}\frac{g_j}{F^\frac 12}+(n-|S|)\hat{g}}+\frac{(n-|S|)\cdot \hat{g}}{\sum_{t \in S}\frac{g_j}{F^\frac 12 }+(n-|S|-1)\hat{g}}\\
&=& \frac{|S|\cdot g_j}{(|S|-1)\frac{g_j}{F^\frac 12}+(n-|S|)\hat{g}}+\frac{(n-|S|)\cdot \hat{g}}{|S|\frac{g_j}{F^\frac 12 }+(n-|S|-1)\hat{g}} \enspace .
\end{eqnarray*}
Let $g = \frac{g_j}{\hat{g}}$, then:
$$H(\G,\g) \geq \frac{|S|\cdot g}{(|S|-1)\frac{g}{F^\frac 12}+n-|S|}+\frac{n-|S|}{|S|\frac{g}{F^\frac 12 }+n-|S|-1} \enspace .$$
If $|S|=0$, $H(\G,\g) \geq \frac{n}{n-1} \geq 1 \geq \sqrt{F}$. Else, due to Lemma \ref{optimize_g}, $H(\G,\g) \geq \sqrt{F}$. 
\end{proof}

%%% Local Variables: 
%%% mode: latex
%%% TeX-master: "allpay"
%%% End: 

%% file: appendixSingleFirst.tex
\section{Conventional Procurement}
\label{firstPriceRevenue}

In this section we give bounds on the expected revenue and maximum bid of the 
single-item first-price auction. In the following, we drop the word expected when
referring to the revenue or to the maximum bid.

\begin{theorem}
\label{revFirstPrice}
In any mixed Nash equilibrium, 
the revenue and the maximum bid lie between the two 
highest valuations. There further exists a tie-breaking rule, 
such that in the worst-case, these quantities match the second highest valuation 
(This can also be achieved, under the no-overbidding assumption).
\end{theorem}

\begin{lemma}
\label{u=0_bid>=v}
In any mixed Nash equilibrium, 
if the expected utility of any player $i$ with valuation $v_i$ is $0$, then with 
probability $1$ the maximum bid is at least $v_i$. 
\end{lemma}

\begin{proof}
Consider any mixed Nash equilibrium $\b \sim \B$ and let $h=\max_i \{b_i\}$ 
be the highest bid; $h$ is a random variable induced by $\B$. 
For the sake of contradiction, assume that $h$ is {\em strictly} less than 
$v_i$ with probability $p>0$. Then, there exists $\varepsilon >0$ 
such that $h < v_i-\varepsilon$ with probability $p$. Consider now 
the deviation of player $i$ to pure strategy $s_i = v_i - \varepsilon$. 
$s_i$ would be the maximum bid with probability $p$ and therefore the utility 
of player $i$ would be at least $p(v_i-(v_i-\varepsilon))=p\cdot \varepsilon >0$. 
This contradicts the fact that $\B$ is an equilibrium and completes the proof of lemma. \qed
\end{proof}

\begin{lemma}
\label{uOfv2=0}
In any mixed Nash equilibrium, if $v$ 
is the highest valuation, any player with valuation strictly 
less than $v$ has expected utility equal to $0$.
\end{lemma}

\begin{proof}
In \cite{CKST13} (Theorem 5.4), they proved that the price of anarchy of mixed Nash equilibria, 
for the single-item first-price auction, is exactly $1$. This means that the player(s)
with the highest valuation gets the item with probability $1$. Therefore, any player 
with valuation strictly less than $v$ gets the item with zero probability and hence, 
her expected utility is $0$. \qed
\end{proof}

Consider the players ordered based on their valuations so that 
$v_1 \geq v_2 \geq v_3 \geq \ldots \geq v_n$. 
In order to prove Theorem \ref{revFirstPrice}, we distinguish between two cases: 
i) $v_1 >v_2$ and ii) $v_1 = v_2$.

\begin{lemma}
\label{v1>v2FirstPrice}
If $v_1>v_2$, the maximum bid of any mixed Nash equilibrium, 
is at least $v_2$ and at most $v_1$. 
If we further assume no-overbidding, the maximum bid is exactly $v_2$.
\end{lemma}

\begin{proof}
If $v_1>v_2$, by Lemma \ref{uOfv2=0}, the expected utility of player $2$ equals 
$0$. From Lemma \ref{u=0_bid>=v}, the highest bid is at least $v_2$ 
with probability $1$. Moreover, if there exist players bidding above $v_1$ 
with positive probability, then at least one of them (whoever gets the item 
with positive probability) would have negative utility for that bid and would prefer 
to deviate to $0$; so, the bidding profile couldn't be an equilibrium. Therefore, 
the maximum bid lies between $v_1$ and $v_2$. 

If we further assume no-overbidding, nobody, apart from player $1$, would 
bid above $v_2$. So, the same hold for player $1$, who has an incentive to bid 
arbitrarily close to $v_2$.  \qed
\end{proof}

\begin{corollary}
If $v_1>v_2$, there exists a tie breaking rule, under which the maximum bid 
of the worst-case mixed Nash equilibrium is exactly $v_2$. 
\end{corollary}

\begin{proof}
Due to Lemma \ref{v1>v2FirstPrice}, it is sufficient to show a tie breaking rule, where 
there exists a mixed Nash equilibrium with highest bid equal to $v_2$. 
Consider the tie-breaking rule where, in a case of a tie with player $1$ (the bidder of the highest valuation), 
the item is always allocated to player $1$. Under this tie-breaking rule, the pure strategy 
profile, where everybody bids $v_2$ is obviously a pure Nash equilibrium, with 
$v_2$ being the maximum bid. \qed
\end{proof}

\begin{lemma}
\label{v1=v2FirstPrice}
If $v_1=v_2$, the maximum bid of any mixed Nash equilibrium, equals $v_2$.
\end{lemma}

\begin{proof}
Consider a set $S$ of $k \geq 2$ players having the same valuation $v_1=v_2=\ldots = v_k = v$ 
and the rest having a valuation strictly less than $v$. 
For any mixed Nash equilibrium $\b \sim \B$ and any player $i$, let $G_i$ and $F_i$ 
be the CDFs of $b_i$ and $\max_{i' \neq i} b_{i'}$, respectively. We define $l_i = \inf\{x|G_i(x) > 0\}$ 
to be the infimum value of player's $i$ support in $\B$. We would like to prove that 
$\max_i l_i = v$. For the sake of contradiction, assume that $\max_i l_i < v$ (Assumption 1).

We next prove that, under Assumption 1, $l_i=l$ for any player $i \in S$ and for some $0 \leq l < v$. 
We will assume that $l_{j} < l_i$ for some players $i,j \in S$ (Assumption 2) and we will 
show that Assumption 2 contradicts Assumption 1. There exists 
$\varepsilon >0$ such that $l_{j}+\varepsilon < l_i$. Moreover, based on the 
definition of $l_{j}$, for any $\varepsilon' >0$, $G_j(l_j+\varepsilon') > 0$ 
and so $G_j(l_j+\varepsilon) > 0$. When player's $j$ bid is derived by the 
interval $[l_j,l_j+\varepsilon]$, she receives the item with zero probability, 
since $l_i > l_{j}+\varepsilon$. Therefore, for any bid of her support that is at most  
$l_{j}+\varepsilon$, her utility is zero ($G_j(l_j+\varepsilon) > 0$, so there should be such a bid). 
Since $\B$ is a mixed Nash equilibrium, her total expected utility should also be zero. 
In that case, Lemma \ref{u=0_bid>=v} contradicts Assumption 1, and therefore Assumption 2 
cannot be true (under Assumption 1). Thus, for any player $i \in S$, $l_i=l$ for some $0 \leq l < v$. 

Moreover, Lemma \ref{uOfv2=0} indicates that no player $i \notin S$ bids above $l$ with positive 
probability, i.e. $G_i(l)=1$ for all $i \notin S$. 
We now show that for any $i \in S$, $G_i$ cannot have a mass point at $l$, i.e. $G_i(l)=0$ for all $i \in S$.\\ 
{\em Case 1.} If $G_i(l) > 0$ for all $i$, then $p = \prod_i G_i(l) > 0$ is the probability that the highest bid is $l$, 
or more precisely, it is the probability that all players in $S$ bid $l$ and a tie occurs. 
Given that this event occurs, there exists a player $j \in S$ that gets the item 
with probability $p_j$ strictly less than $1$ (this is the conditional probability). Therefore, player $j$ 
has an incentive to deviate from $l$ to $l+\varepsilon$, for $\varepsilon < (1-p_j)(v-l)$ (so that 
$p_j(v-l)< v-(l+\varepsilon)$); this contradicts the fact that $\B$ is an equilibrium. \\
{\em Case 2.} If $G_i(l) > 0$ and $G_j(l)=0$ for some $i,j \in S$, then $l$ is in the support of 
player $i$, but she does never receives the item when she bids $l$, since player $j$ bids above $l$ 
with probability $1$. Therefore, the expected utility of player $i$ is $0$ and due to Lemma \ref{u=0_bid>=v} 
this cannot happen under Assumption 1. 

Overall, we have proved so far that, under Assumption 1 (that now has become $l < v$), $G_i(l)=0$ for all $i \in S$ and $G_i(l)=1$ 
for all $i \notin S$. Since $k \geq 2$, $F_i(l)=\prod_{i'\neq i} G_{i'}(l) = 0$ for all $i$. 
Consider any player $i\in S$ and let $u_i$ be her expected utility. Based on the definition of $l_i$, 
for any $\varepsilon >0$, there exists $x(\varepsilon) \in [l,l+\varepsilon]$, such that $x(\varepsilon)$ is in the support of 
player $i$. Therefore, $u_i \leq F_i(x(\varepsilon))(v-x(\varepsilon)) \leq F_i(l+\varepsilon)(v-l)$. 
As $F_i$ is a CDF, it should be right-continuous and so for any $\delta > 0$, there exists some 
$\varepsilon > 0$, such that $F_i(l+\varepsilon)(v-l) < \delta$ and therefore, $u_i < \delta$. We can contradict Assumption 1, right away by 
using Lemma \ref{u=0_bid>=v}, but we give a bit more explanation. 
Assume that, in $\B$, the maximum bid $h$ is strictly 
less than $v$ with probability $p>0$. Then, there exists some $\varepsilon' >0$, such that 
$h < v-\varepsilon'$ with probability $p$. If we consider any $\delta < p(v-\varepsilon')$, 
it is straight forward to see that player $i$ has an incentive to deviate to the pure 
strategy $v-\varepsilon'$. Therefore, we showed that Assumption 1 cannot hold and so the 
highest bid is at least $v$ with probability $1$. Similar to the proof of Lemma \ref{v1>v2FirstPrice}, 
nobody will bid above $v$ in any mixed Nash equilibrium. 
\end{proof}

%%% Local Variables:
%%% mode: latex
%%% TeX-master: "allpay"
%%% End:

%% file: allpay.bbl
\begin{thebibliography}{10}

\bibitem{BKV96}
Michael~R. Baye, Dan Kovenock, and Casper~G. de~Vries.
\newblock The all-pay auction with complete information.
\newblock {\em Economic Theory}, 8(2):291--305, August 1996.

\bibitem{BR11}
Kshipra Bhawalkar and Tim Roughgarden.
\newblock {Welfare guarantees for combinatorial auctions with item bidding}.
\newblock In {\em SODA '11}. SIAM, January 2011.

\bibitem{Bik99}
Sushil Bikhchandani.
\newblock {Auctions of Heterogeneous Objects}.
\newblock {\em Games and Economic Behavior}, January 1999.

\bibitem{CDW12}
Yang Cai, Constantinos Daskalakis, and S.~Matthew Weinberg.
\newblock An algorithmic characterization of multi-dimensional mechanisms.
\newblock In {\em Proceedings of the Forty-fourth Annual ACM Symposium on
  Theory of Computing}, STOC '12, pages 459--478, New York, NY, USA, 2012. ACM.

\bibitem{CV14}
Ioannis Caragiannis and Alexandros~A. Voudouris.
\newblock Welfare guarantees for proportional allocations.
\newblock {\em SAGT '14}, 2014.

\bibitem{CHS12}
Shuchi Chawla, Jason~D. Hartline, and Balasubramanian Sivan.
\newblock Optimal crowdsourcing contests.
\newblock In {\em SODA 2012, Kyoto, Japan, January 17-19, 2012}, pages
  856--868, 2012.

\bibitem{CKS08}
George Christodoulou, Annam{\'a}ria Kov{\'a}cs, and Michael Schapira.
\newblock {Bayesian Combinatorial Auctions}.
\newblock In {\em ICALP '08}. Springer-Verlag, July 2008.

\bibitem{CKST13}
George Christodoulou, Annam{\'{a}}ria Kov{\'{a}}cs, Alkmini Sgouritsa, and
  Bo~Tang.
\newblock Tight bounds for the price of anarchy of simultaneous first price
  auctions.
\newblock {\em ACM Transactions on Economics and Computation (to appear)},
  2015.

\bibitem{KMST13}
Bart de~Keijzer, Evangelos Markakis, Guido Sch{\"a}fer, and Orestis Telelis.
\newblock {On the Inefficiency of Standard Multi-Unit Auctions}.
\newblock In {\em ESA'13}, March 2013.

\bibitem{DV09}
Dominic DiPalantino and Milan Vojnovic.
\newblock Crowdsourcing and all-pay auctions.
\newblock In {\em EC '09}, pages 119--128, New York, NY, USA, 2009. ACM.

\bibitem{DFK11}
Shahar Dobzinski, Hu~Fu, and Robert~D. Kleinberg.
\newblock Optimal auctions with correlated bidders are easy.
\newblock In {\em Proceedings of the Forty-third Annual ACM Symposium on Theory
  of Computing}, STOC '11, pages 129--138, New York, NY, USA, 2011. ACM.

\bibitem{FFGL13}
Michal Feldman, Hu~Fu, Nick Gravin, and Brendan Lucier.
\newblock {Simultaneous Auctions are (almost) Efficient}.
\newblock In {\em STOC '13}, September 2013.

\bibitem{HKMN11}
Avinatan Hassidim, Haim Kaplan, Yishay Mansour, and Noam Nisan.
\newblock {Non-price equilibria in markets of discrete goods}.
\newblock In {\em EC '11}. ACM, June 2011.

\bibitem{johari_efficiency_2004}
Ramesh Johari and John~N Tsitsiklis.
\newblock Efficiency loss in a network resource allocation game.
\newblock {\em Mathematics of Operations Research}, 29(3):407–435, August
  2004.

\bibitem{Kel97}
Frank Kelly.
\newblock Charging and rate control for elastic traffic.
\newblock {\em Eur. Trans. Telecomm.}, 8(1):33–37, January 1997.

\bibitem{KP99}
Elias Koutsoupias and Christos Papadimitriou.
\newblock {Worst-case equilibria}.
\newblock In {\em STACS '99}. Springer-Verlag, March 1999.

\bibitem{Kri02}
Vijay Krishna.
\newblock {\em Auction Theory}.
\newblock Academic Press, 2002.

\bibitem{MT12}
Evangelos Markakis and Orestis Telelis.
\newblock Uniform price auctions: Equilibria and efficiency.
\newblock In {\em SAGT}, pages 227--238, 2012.

\bibitem{MS96}
Dov Monderer and Lloyd~S. Shapley.
\newblock Potential games.
\newblock {\em Games and Economic Behavior}, 14(1):124 -- 143, 1996.

\bibitem{Rou14}
Tim Roughgarden.
\newblock Barriers to near-optimal equilibria.
\newblock In {\em FOCS 2014, Philadelphia, PA, USA, October 18-21, 2014}, pages
  71--80, 2014.

\bibitem{Sie03}
Ron Siegel.
\newblock All-pay contests.
\newblock {\em Econometrica}, 77(1):71--92, January 2009.

\bibitem{simon1990discontinuous}
Leo~K Simon and William~R Zame.
\newblock Discontinuous games and endogenous sharing rules.
\newblock {\em Econometrica: Journal of the Econometric Society}, pages
  861--872, 1990.

\bibitem{ST13}
Vasilis Syrgkanis and Eva Tardos.
\newblock {Composable and Efficient Mechanisms}.
\newblock In {\em STOC '13: Proceedings of the 45th symposium on Theory of
  Computing}, November 2013.

\bibitem{Vic61}
William Vickrey.
\newblock Counterspeculation, auctions, and competitive sealed tenders.
\newblock {\em The Journal of finance}, 16(1):8--37, 1961.

\end{thebibliography}
